\documentclass[apsrev4-1,preprint,prd,showpacs,floatfix,nofootinbib,superscriptaddress]{revtex4-1}
\usepackage{epsf,color,colordvi}
\usepackage[utf8]{inputenc}
\usepackage{caption}
\usepackage{graphicx}
\usepackage{amsmath}
\usepackage{times}
\usepackage{subfig}
\usepackage{ulem}  % This package redefines {} to underline
\usepackage{placeins}

\begin{document}

\title{New physics in neutral $\textbf{D}\rightarrow\textbf{VV}$ modes}

\author{Aritra Biswas}
\email{aritrab@imsc.res.in}
\affiliation{The Institute of Mathematical Sciences, Chennai 600113, India}
\author{Sanjoy Mandal}
\email{smandal@imsc.res.in}
\affiliation{The Institute of Mathematical Sciences, Chennai 600113, India}

\begin{abstract}
We show that it is possible to construct observables to test the existence of new physics in a model independent way for the $D^0\rightarrow VV$ modes using a time-dependent analysis of the neutral $D$ meson. We show that it is possible to identify whether the NP is due to decay, mixing or a combination of both. We also provide numerical estimates for the polarization amplitudes for the $D^0\rightarrow\overline{K^{*0}}\rho^0$ mode and show that our analysis is consistent with the present data.
\end{abstract}

%\pacs{}

\date{\today} 

\maketitle

\section{Introduction}
With the discovery of the Higgs Boson in 2012 by both the ATLAS~\cite{Higgs_atlas} and the CMS~\cite{Higgs_cms} collaborations, our understanding of the physics of the fundamental particles up to the $1$ TeV scale, otherwise known as the Standard Model (SM), has been experimentally confirmed. However, though we know with certainty that SM is the correct picture, we also know with an equal degree of certainty that it is not the complete one. The mathematical framework of the SM is inconsistent with that of general relativity, but the later has been experimentally verified. Other experimental discoveries such as the neutrino oscillations, along with theoretical questions such as the origin of mass, the matter antimatter asymmetry, the dark matter and dark energy, the strong CP problem etc provide ample evidence for the existence of new physics (NP) beyond the SM.

Over the past three decades, flavour physics has emerged as an important testing ground for the existence of NP. For example, tensions between SM expectations and experimental results have been found in B physics for observables such as the isospin asymmetry $A_I(B\rightarrow K\mu^+\mu^-)$~\cite{isospin}, the longitudinal polarization fraction in $B_s\rightarrow K^*K^*$~\cite{long_pol_1, long_pol_2}, $R_{D^*}^{\tau,l}=\frac{\text{BR}(B\rightarrow D^*\tau\nu_{\tau})}{\text{BR}(B\rightarrow D^*l\nu_l)}$, $R_{D}^{\tau,l}=\frac{\text{BR}(B\rightarrow D\tau\nu_{\tau})}{\text{BR}(B\rightarrow Dl\nu_l)}$~\cite{rddst_1, rddst_2, rddst_3, rddst_4, rddst_5, rddst_6, rddst_7} and $R_K=\frac{\text{BR}(B\rightarrow K\mu^+\mu^-)}{\text{BR}(B\rightarrow Ke^+e^-)}$~\cite{b_kll_1, b_kll_2}. A number of theoretical works have been and are still being undertaken following these results. Top quark physics seems to be important for the search of NP in the up quark sector. However, no signal for NP has yet been detected in the top quark sector.

Searches for NP have also been carried out in charm. A considerable amount of work on the $D\rightarrow PP$ and $D\rightarrow VP$ modes~\cite{pp_pv_1, pp_pv_2, pp_pv_3, pp_pv_4, pp_pv_5, pp_pv_6, pp_pv_7, pp_pv_8, pp_pv_9, pp_pv_10, pp_pv_11, pp_pv_12, pp_pv_13, pp_pv_14, pp_pv_15, pp_pv_16, pp_pv_17, pp_pv_18, pp_pv_19, pp_pv_20, pp_pv_21, pp_pv_22, vv_1} has been undertaken over the last thirty years. In 2012, LHCb~\cite{LHCb_charm} and CDF~\cite{CDF_charm} reported the first observation of a CP asymmetry between the $D^0\rightarrow\pi^+\pi^-$ and the $D^0\rightarrow K^+K^-$ modes. This was followed by a large amount of work~\cite{np_1, np_2, np_3, np_4, np_5, np_6, np_7, np_8, np_9, np_10, np_11, np_12, np_13, np_14, np_15, np_16, np_17}, where the authors  mostly used NP models to explain the same. The $3.2\sigma$ hint has since then slowly disappeared. However, the $D\rightarrow VV$ hadronic modes have received lesser attention~\cite{vv_1, vv_2}.

It is not an easy task to have a solid theoretical understanding of all the charm hadronic modes. This is due to the mass of the charm quark, which, unlike the bottom quark is not sufficiently heavy for the realization of the infinitely heavy quark limit. Hence, the well known approaches based on QCD that lead to satisfactory predictions for $B$ decays like the heavy quark effective theory~\cite{hqet_1, hqet_2}, the QCD-factorization~\cite{qfac_1, qfac_2}, the perturbative QCD approach~\cite{qper_1, qper_2, qper_3, qper_4} and the soft-collinear effective theory~\cite{scet}, fail to achieve the same for $D$ decays. The charm quark is also not light enough for the application of a chiral perturbation theory. Furthermore, in the case of hadronic vector final states, the calculation of the form factors poses greater difficulty than their $D\rightarrow P$ counterparts.

In absence of any reliable and effective theoretical models, it might be a good idea to look for possible NP in $D\rightarrow VV$ decays in a model independent way. This is what we have tried to achieve in this paper. The method is similar to the one used in~\cite{ang_an_B} for $B$ decays. The idea is to define observables which can be experimentally measured, linear combinations of which have a value inconsistent with zero under the presence of new physics. Although extensively used for the $B\rightarrow VV$ decays, such a formalism, to the best of our knowledge, has not yet been pursued for the $D\rightarrow VV$ hadronic decays. 

Our paper is organised as follows. In section~\ref{ang_an} we describe the formalism for defining such model independent observables in detail. The next section (section~\ref{pol}) is about the connection between the polarization basis that we use for our analysis and the $a$, $b$, $c$ amplitudes used in the most general covariant expression for a heavy pseudoscalar to two-body vector decay amplitude. We then discuss our observables and the effect NP has on them in section~\ref{results}. We also extract numerical values for the amplitudes in the polarization basis for $D^0\rightarrow \overline{K^{*0}}\rho^0$ mode, which is the best measured $D\rightarrow VV$ mode in~\cite{pdg}. Finally, we summarize and conclude in section~\ref{summary}.

\section{The time-dependent analysis\label{ang_an}}
Consider a $D\rightarrow V_1V_2$ decay. Let the SM contribution to this decay be parametrized by a single decay amplitude along with a corresponding strong phase. The NP contribution will in general be characterised by a different decay amplitude with a different strong phase, along with a NP weak phase. The corresponding CP conjugate decay will then have the same components, with the sign of the NP weak phase reversed. A decay of the type $D\rightarrow\bar{V_1}\bar{V_2}$, where the bar denotes the CP conjugate state, and it's corresponding CP conjugate decay will in general be parametrized by some different SM and NP parameters. Hence, for each of the above cases, the decay amplitude for each of the three possible helicity states may be written as:
\begin{eqnarray}\label{amplitude}
 A_{\lambda}&\equiv& \text{Amp}(D\rightarrow V_1V_2)=a_{\lambda}e^{i\delta_{\lambda}^a}+b_{\lambda}e^{i\phi}e^{i\delta_{\lambda}^b}\nonumber\\
 \bar{A}_{\bar\lambda}&\equiv& \text{Amp}(\bar{D}\rightarrow \bar{V_1}\bar{V_2})=a_{\lambda}e^{i\delta_{\lambda}^a}+b_{\lambda}e^{-i\phi}e^{i\delta_{\lambda}^b}\nonumber\\
 A_{\bar\lambda}&\equiv& \text{Amp}(D\rightarrow \bar{V_1}\bar{V_2})=c_{\lambda}e^{i\delta_{\lambda}^c}+d_{\lambda}e^{i\gamma}e^{i\delta_{\lambda}^d}\nonumber\\
 \bar{A}_{\lambda}&\equiv& \text{Amp}(\bar{D}\rightarrow V_1V_2)=c_{\lambda}e^{i\delta_{\lambda}^c}+d_{\lambda}e^{-i\gamma}e^{i\delta_{\lambda}^d}.
\end{eqnarray}
In the above, $a_{\lambda}$, $c_{\lambda}$ represent the SM decay amplitudes; $\delta_{\lambda}^a$ and $\delta_{\lambda}^c$ denote the SM strong phases; $b_{\lambda}$, $d_{\lambda}$ are the NP decay amplitudes;  $\delta_{\lambda}^b$ and $\delta_{\lambda}^d$ the NP strong phases and $\phi$, $\gamma$ are the NP weak phases. The helicity index $\lambda$ takes the values ${0, \parallel, \perp}$ denoting
the longitudinal, transverse parallel and the transverse perpendicular helicities respectively. Using CPT invariance, the full amplitude for each of the above decays can be written as
\begin{eqnarray}\label{amplitude}
 \mathcal{A}_{f}&\equiv&\text{Amp}\left(D(t)\rightarrow V_{1}V_{2}\right)={A_{f}}_{0}g_{0}+{A_{f}}_{\parallel}g_{\parallel}+i{A_{f}}_{\perp}g_{\perp}\nonumber\\
  \mathcal{A}_{\bar f}&\equiv&\text{Amp}\left(D(t)\rightarrow\overline{V}_{1}\overline{V}_{2}\right)={A_{\bar f}}_{0}g_{0}+{A_{\bar f}}_{\parallel}g_{\parallel}+i{A_{\bar f}}_{\perp}g_{\perp}\nonumber\\
 \bar{\mathcal{A}}_{f}&\equiv&\text{Amp}\left(\overline{D}(t)\rightarrow V_{1}V_{2}\right)={\overline{A}_{f}}_{0}g_{0}+{\overline{A}_{f}}_{\parallel}g_{\parallel}-i{\overline{A}_{f}}_{\perp}g_{\perp}\nonumber\\
 \bar{\mathcal{A}}_{\bar f}&\equiv&\text{Amp}\left(\overline{D}(t)\rightarrow\overline{V}_{1}\overline{V}_{2}\right)={\overline{A}_{\bar f}}_{0}g_{0}+
 {\overline{A}_{\bar f}}_{\parallel}g_{\parallel}-i{\overline{A}_{\bar f}}_{\perp}g_{\perp}.
\end{eqnarray}
where $g_{\lambda}$'s are basically functions of the angles describing the kinematics for the 
the corresponding decay. 
Based on the parametrization discussed above, one can now look into the time dependent decay rates of a neutral $D$ meson going to two vector final states and perform an angular analysis of all the $D(t)\to V_1V_2$, $D(t)\to\overline{V}_1\overline{V}_2$ modes and their CP conjugate processes. The general expressions for the time-dependent decay rates of a neutral meson $M^0$ going to two vector final states are given by
\begin{eqnarray}
 \Gamma  \left(M^0(t)\to V_1V_2\right)=&N_fe^{-\text{$\Gamma $t}}\left(\left|\mathcal{A}_f\right|^2+\left|\frac{q}{p} \bar{\mathcal{A}}_f\right|^2\right)\cosh\left(\text{y$\Gamma $t}\right)
 +\left(\left|\mathcal{A}_f\right|^2-\left|\frac{q}{p} \bar{\mathcal{A}}_f\right|^2\right)\cos\left(\text{x$\Gamma $t}\right)\nonumber&\\
 &+2\text{Re}\left(\left(\frac{q}{p}\right)\mathcal{A}_{f}^{*}\bar{\mathcal{A}}_{f}\right)\sinh(\text{y$\Gamma$t})
 -2\text{Im}\left(\left(\frac{q}{p}\right)\mathcal{A}_{f}^{*}\bar{\mathcal{A}}_{f}\right)\text{sin}\left(\text{x$\Gamma$t}\right)\nonumber&\\
  \Gamma  \left(M^0(t)\to \overline{V}_1 \overline{V}_2 \right)=&N_fe^{-\text{$\Gamma $t}}\left(\left|\mathcal{A}_{\bar f}\right|^2+\left|\frac{q}{p} \bar{\mathcal{A}}_{\bar f}\right|^2\right)\cosh\left(\text{y$\Gamma $t}\right)
 +\left(\left|\mathcal{A}_{\bar f}\right|^2-\left|\frac{q}{p} \bar{\mathcal{A}}_{\bar f}\right|^2\right)\cos\left(\text{x$\Gamma $t}\right)\nonumber&\\
 &+2\text{Re}\left(\left(\frac{q}{p}\right)\mathcal{A}_{\bar f}^{*}\bar{\mathcal{A}}_{\bar f}\right)\sinh(\text{y$\Gamma$t})
 -2\text{Im}\left(\left(\frac{q}{p}\right)\mathcal{A}_{\bar f}^{*}\bar{\mathcal{A}}_{\bar f}\right)\text{sin}\left(\text{x$\Gamma$t}\right)\nonumber&\\
  \Gamma  \left(\overline{M^0}(t)\to V_1V_2\right)=&N_fe^{-\text{$\Gamma $t}}\left(\left|\frac{p}{q} \mathcal{A}_f\right|^2+\left|\bar{\mathcal{A}}_f\right|^2\right)\cosh\left(\text{y$\Gamma $t}\right)
 -\left(\left|\frac{p}{q} \mathcal{A}_f\right|^2-\left|\bar{\mathcal{A}}_f\right|^2\right)\cos\left(\text{x$\Gamma $t}\right)\nonumber&\\
 &+2\text{Re}\left(\left(\frac{p}{q}\right)\mathcal{A}_{f}\bar{\mathcal{A}}_{f}^{*}\right)\sinh(\text{y$\Gamma$t})
 -2\text{Im}\left(\left(\frac{p}{q}\right)\mathcal{A}_{f}\bar{\mathcal{A}}_{f}^{*}\right)\text{sin}\left(\text{x$\Gamma$t}\right)\nonumber&\\
  \Gamma  \left(\overline{M^0}(t)\to \overline{V}_1 \overline{V}_2\right)=&N_fe^{-\text{$\Gamma $t}}\left(\left|\frac{p}{q} \mathcal{A}_{\bar f}\right|^2+\left|\bar{\mathcal{A}}_{\bar f}\right|^2\right)\cosh\left(\text{y$\Gamma $t}\right)
 -\left(\left|\frac{p}{q} \mathcal{A}_{\bar f}\right|^2-\left|\bar{\mathcal{A}}_{\bar f}\right|^2\right)\cos\left(\text{x$\Gamma $t}\right)\nonumber&\\
 &+2\text{Re}\left(\left(\frac{p}{q}\right)\mathcal{A}_{\bar f}\bar{\mathcal{A}}_{\bar f}^{*}\right)\sinh(\text{y$\Gamma$t})
 -2\text{Im}\left(\left(\frac{p}{q}\right)\mathcal{A}_{\bar f}\bar{\mathcal{A}}_{\bar f}^{*}\right)\text{sin}\left(\text{x$\Gamma$t}\right).&
\end{eqnarray}
Here $x=\frac{\Delta m}{\Gamma}$, $y=\frac{\Delta\Gamma}{2\Gamma}$, $\Delta m=m_{H}-m_{L}$, $\Delta\Gamma=\Gamma_{H}-\Gamma_{L}$. The indices H, L stand for the heavy and light
mass eigenstates. $N_{f}$ is a time independent normalization factor. Decays occurring without $M^0-\overline{M^0}$ oscillations (pure decays) are given by the terms proportional to 
$\left|\mathcal{A}_{f}\right|^{2}$ or $\left|\bar{A}_{f}\right|^{2}$. Terms proportional to $\left|\frac{q}{p}\bar{\mathcal{A}}_{f}\right|^{2}$ or
$\left|\frac{p}{q}\mathcal{A}_{f}\right|^{2}$ are due to decays following the $M^0-\overline{M^0}$ oscillation (decays after mixing).

BABAR, while measuring the effect of CP-violation on the time-dependent decay rates in the $D\rightarrow K\pi$ system, parametrized the decay rates in powers of $\Gamma t$~\cite{BABAR}. They truncate the series at second order in $\Gamma t$. However the sensitivity of the coefficient of the second order term to the decay rate is less than that of the coefficient of the first order term~\cite{Ball}. We follow the same parametrization and keep the terms up to the quadratic order in $\Gamma t$.
It is clear that our observables will be linear combinations of the coefficients of the sinh, sin, cosh and cos terms. For the case of neutral charm decays, the corresponding decay rates look like
\begin{align}\label{decay rate}
\Gamma  \left(D^0(t)\to V_{1}V_{2}\right)=&N_fe^{-\text{$\Gamma $t}}\bigg(2\left|\mathcal{A}_{f}\right|^{2}+\left[2yRe\left(\frac{q}{p}\mathcal{A}_{f}^{*}\bar{\mathcal{A}}_{f}\right)
-2xIm\left(\frac{q}{p}\mathcal{A}_{f}^{*}\bar{\mathcal{A}}_{f}\right)\right]\Gamma t\nonumber\\
&+\left[\frac{(y^{2}-x^{2})}{2}\left|\mathcal{A}_{f}\right|^{2}+\frac{(x^{2}+y^{2})}{2}\left|\frac{q}{p}\bar{\mathcal{A}}_{f}\right|^{2}\right](\Gamma t)^{2}\bigg)\nonumber\\
\Gamma  \left(D^0(t)\to \overline{V}_{1} \overline{V}_{2}\right)=&N_fe^{-\text{$\Gamma $t}}\bigg(2\left|\mathcal{A}_{\bar f}\right|^{2}+\left[2yRe\left(\frac{q}{p}\mathcal{A}_{\bar f}^{*}\bar{\mathcal{A}}_{\bar f}\right)
-2xIm\left(\frac{q}{p}\mathcal{A}_{\bar f}^{*}\bar{\mathcal{A}}_{\bar f}\right)\right]\Gamma t\nonumber\\
&+\left[\frac{(y^{2}-x^{2})}{2}\left|\mathcal{A}_{\bar f}\right|^{2}+\frac{(x^{2}+y^{2})}{2}\left|\frac{q}{p}\bar{\mathcal{A}}_{\bar f}\right|^{2}\right](\Gamma t)^{2}\bigg)\nonumber\\
\Gamma  \left(\overline{D^0}(t)\to V_{1}V_{2}\right)=&N_fe^{-\text{$\Gamma $t}}\bigg(2\left|\bar{\mathcal{A}}_{f}\right|^{2}+\left[2yRe\left(\frac{p}{q}\bar{\mathcal{A}}_{f}^{*}\mathcal{A}_{f}\right)
-2xIm\left(\frac{p}{q}\bar{\mathcal{A}}_{f}^{*}\mathcal{A}_{f}\right)\right]\Gamma t\nonumber\\
&+\left[\frac{(y^{2}-x^{2})}{2}\left|\bar{\mathcal{A}}_{f}\right|^{2}+\frac{(x^{2}+y^{2})}{2}\left|\frac{p}{q}\mathcal{A}_{f}\right|^{2}\right](\Gamma t)^{2}\bigg)\nonumber\\
\Gamma  \left(\overline{D^0}(t)\to\overline{V}_{1}\overline{V}_{2}\right)=&N_fe^{-\text{$\Gamma $t}}\bigg(2\left|\bar{\mathcal{A}}_{\bar f}\right|^{2}+\left[2yRe\left(\frac{p}{q}\bar{\mathcal{A}}_{\bar f}^{*}\mathcal{A}_{\bar f}\right)
-2xIm\left(\frac{p}{q}\bar{\mathcal{A}}_{\bar f}^{*}\mathcal{A}_{\bar f}\right)\right]\Gamma t\nonumber\\
&+\left[\frac{(y^{2}-x^{2})}{2}\left|\bar{\mathcal{A}}_{\bar f}\right|^{2}+\frac{(x^{2}+y^{2})}{2}\left|\frac{p}{q}\mathcal{A}_{\bar f}\right|^{2}\right](\Gamma t)^{2}\bigg)
\end{align}

Using eqns~(\ref{amplitude}) and~(\ref{decay rate}) we can write the time dependent decay rates as
\begin{eqnarray}
 \Gamma\left(D^0(t)\to V_{1}V_{2}\right)&=&e^{-\Gamma t}\sum_{\lambda\leq\lambda'}\left(X^{V_{1}V_{2}}_{\lambda\lambda'}+Y^{V_{1}V_{2}}_{\lambda\lambda'}\,\Gamma t+Z^{V_{1}V_{2}}_{\lambda\lambda'}\,(\Gamma t)^2\right)g_{\lambda}g_{\lambda'}\nonumber\\
 \Gamma\left(D^0(t)\to\overline{V}_{1}\overline{V}_{2}\right)&=&e^{-\Gamma t}\sum_{\lambda\leq\lambda'}\left(X^{\overline{V}_{1}\overline{V}_{2}}_{\lambda\lambda'}+Y^{\overline{V}_{1}\overline{V}_{2}}_{\lambda\lambda'}\,\Gamma t+Z^{\overline{V}_{1}\overline{V}_{2}}_{\lambda\lambda'}\,(\Gamma t)^2\right)g_{\lambda}g_{\lambda'}\nonumber\\
 \Gamma\left(\overline{D^0}(t)\to V_{1}V_{2}\right)&=&e^{-\Gamma t}\sum_{\lambda\leq\lambda'}\left(\bar{X}^{V_{1}V_{2}}_{\lambda\lambda'}+\bar{Y}^{V_{1}V_{2}}_{\lambda\lambda'}\,\Gamma t+\bar{Z}^{V_{1}V_{2}}_{\lambda\lambda'}\,(\Gamma t)^2\right)g_{\lambda}g_{\lambda'}\nonumber\\
 \Gamma\left(\overline{D^0}(t)\to\overline{V}_{1}\overline{V}_{2}\right)&=&e^{-\Gamma t}\sum_{\lambda\leq\lambda'}\left(\bar{X}^{\overline{V}_{1}\overline{V}_{2}}_{\lambda\lambda'}+\bar{Y}^{\overline{V}_{1}\overline{V}_{2}}_{\lambda\lambda'}\,\Gamma t+\bar{Z}^{\overline{V}_{1}\overline{V}_{2}}_{\lambda\lambda'}\,(\Gamma t)^2\right)g_{\lambda}g_{\lambda'}.
 \end{eqnarray}
Thus by a time dependent angular analysis of the decay modes $D^0(t)\to V_{1}V_{2}$, $D^0(t)\to\overline{V}_{1}\overline{V}_{2}$,
$\overline{D^0}(t)\to V_{1}V_{2}$,
$\overline{D^0}(t)\to\overline{V}_{1}\overline{V}_{2}$, one can define 72 observables. However, not all of these observables are independent. For example, it can be easily verified that the $Z$'s (the coefficient of the $(\Gamma t)^2$) terms) are linear combinations of the $X$'s. In fact, observables extracted from the coefficients of the terms that are even powers i $\Gamma t$ can always be written as linear combinations of the $X$'s, while those extracted from the coefficients of the terms that are odd powers in $\Gamma t$ can be written as linear combinations of $Y$'s. Hence, in our analysis, we deal with a total of 48 observables (24 $X$'s and 24 $Y$'s). In terms of the amplitudes, these are given by:
\begin{align}\label{pol_obs}
X^{V_{1}V_{2}}_{\lambda\lambda}=&2(A_{f}^{*})_{\lambda}(A_{f})_{\lambda}\nonumber\\
X^{V_{1}V_{2}}_{i\perp}=&-4Im\left[(A_{f}^{*})_{i}(A_{f})_{\perp}\right]\nonumber\\
X^{V_{1}V_{2}}_{0\parallel}=&4Re\left[(A_{f})_{0}^{*}(A_{f})_{\parallel}\right]\nonumber\\
X^{\overline{V}_{1}\overline{V}_{2}}_{\lambda\lambda}=&2(A_{\bar f}^{*})_{\lambda}(A_{\bar f})_{\lambda}\nonumber\\
X^{\overline{V}_{1}\overline{V}_{2}}_{i\perp}=&-4Im\left[(A_{\bar f}^{*})_{i}(A_{\bar f})_{\perp}\right]\nonumber\\
X^{\overline{V}_{1}\overline{V}_{2}}_{0\parallel}=&4Re\left[(A_{\bar f})_{0}^{*}(A_{\bar f})_{\parallel}\right]\nonumber\\
\bar{X}^{V_{1}V_{2}}_{\lambda\lambda}=&2(\bar{A}_{f}^{*})_{\lambda}(\bar{A}_{f})_{\lambda}\nonumber\\
\bar{X}^{V_{1}V_{2}}_{i\perp}=&-4Im\left[(\bar{A}_{f}^{*})_{\perp}(\bar{A}_{f})_{i}\right]\nonumber\\
\bar{X}^{V_{1}V_{2}}_{0\parallel}=&4Re\left[(\bar{A}_{f})_{0}^{*}(\bar{A}_{f})_{\parallel}\right]\nonumber\\
\bar{X}^{\overline{V}_{1}\overline{V}_{2}}_{\lambda\lambda}=&2(\bar{A}_{\bar f}^{*})_{\lambda}(\bar{A}_{\bar f})_{\lambda}\nonumber\\
\bar{X}^{\overline{V}_{1}\overline{V}_{2}}_{i\perp}=&-4Im\left[(\bar{A}_{\bar f}^{*})_{\perp}(\bar{A}_{\bar f})_{i}\right]\nonumber\\
\bar{X}^{\overline{V}_{1}\overline{V}_{2}}_{0\parallel}=&4Re\left[(\bar{A}_{\bar f})_{0}^{*}(\bar{A}_{\bar f})_{\parallel}\right]\nonumber\\
 Y^{V_{1}V_{2}}_{ii}=&2yRe\left[\frac{q}{p}(A_{f}^{*})_{i}(\bar{A}_{f})_{i}\right]
-2xIm\left[\frac{q}{p}(A_{f}^{*})_{i}(\bar{A}_{f})_{i}\right]\nonumber\\
Y^{V_{1}V_{2}}_{\perp\perp}=&-2yRe\left[\frac{q}{p}(A_{f}^{*})_{\perp}(\bar{A}_{f})_{\perp}\right]
+2xIm\left[\frac{q}{p}(A_{f}^{*})_{\perp}(\bar{A}_{f})_{\perp}\right]\nonumber\\
Y^{V_{1}V_{2}}_{i\perp}=&-2yRe\left[\frac{q}{p}i\left((A_{f}^{*})_{i}(\bar{A}_{f})_{\perp}+(A_{f}^{*})_{\perp}(\bar{A}_{f})_{i}\right)\right]
+2xIm\left[\frac{q}{p}i\left((A_{f}^{*})_{i}(\bar{A}_{f})_{\perp}+(A_{f}^{*})_{\perp}(\bar{A}_{f})_{i}\right)\right]\nonumber\\
Y^{V_{1}V_{2}}_{0\parallel}=&2yRe\left[\frac{q}{p}\left((A_{f})_{0}^{*}(\bar{A}_{f})_{\parallel}+(A_{f})_{\parallel}^{*}(\bar{A}_{f})_{0}\right)\right]
-2xIm\left[\frac{q}{p}\left((A_{f})_{0}^{*}(\bar{A}_{f})_{\parallel}+(A_{f})_{\parallel}^{*}(\bar{A}_{f})_{0}\right)\right]\nonumber\\
Y^{\overline{V}_{1}\overline{V}_{2}}_{ii}=&2yRe\left[\frac{q}{p}(A_{\bar f}^{*})_{i}(\bar{A}_{\bar f})_{i}\right]
-2xIm\left[\frac{q}{p}(A_{\bar f}^{*})_{i}(\bar{A}_{\bar f})_{i}\right]\nonumber\\
Y^{\overline{V}_{1}\overline{V}_{2}}_{\perp\perp}=&-2yRe\left[\frac{q}{p}(A_{\bar f}^{*})_{\perp}(\bar{A}_{\bar f})_{\perp}\right]
+2xIm\left[\frac{q}{p}(A_{\bar f}^{*})_{\perp}(\bar{A}_{\bar f})_{\perp}\right]\nonumber\\
Y^{\overline{V}_{1}\overline{V}_{2}}_{i\perp}=&-2yRe\left[\frac{q}{p}i\left((A_{\bar f}^{*})_{i}(\bar{A}_{\bar f})_{\perp}+(A_{\bar f}^{*})_{\perp}(\bar{A}_{\bar f})_{i}\right)\right]
+2xIm\left[\frac{q}{p}i\left((A_{\bar f}^{*})_{i}(\bar{A}_{\bar f})_{\perp}+(A_{\bar f}^{*})_{\perp}(\bar{A}_{\bar f})_{i}\right)\right]\nonumber\\
\end{align}
\begin{align}
Y^{\overline{V}_{1}\overline{V}_{2}}_{0\parallel}=&2yRe\left[\frac{q}{p}\left((A_{\bar f})_{0}^{*}(\bar{A}_{\bar f})_{\parallel}+(A_{\bar f})_{\parallel}^{*}(\bar{A}_{\bar f})_{0}\right)\right]
-2xIm\left[\frac{q}{p}\left((A_{\bar f})_{0}^{*}(\bar{A}_{\bar f})_{\parallel}+(A_{\bar f})_{\parallel}^{*}(\bar{A}_{\bar f})_{0}\right)\right]\nonumber\\
\bar{Y}^{V_{1}V_{2}}_{ii}=&2yRe\left[\frac{p}{q}(\bar{A}_{f}^{*})_{i}(A_{f})_{i}\right]
-2xIm\left[\frac{p}{q}(\bar{A}_{f}^{*})_{i}(A_{f})_{i}\right]\nonumber\\
\bar{Y}^{V_{1}V_{2}}_{\perp\perp}=&-2yRe\left[\frac{p}{q}(\bar{A}_{f}^{*})_{i}(A_{f})_{i}\right]
+2xIm\left[\frac{p}{q}(\bar{A}_{f}^{*})_{i}(A_{f})_{i}\right]\nonumber\\
\bar{Y}^{V_{1}V_{2}}_{i\perp}=&2yRe\left[\frac{p}{q}i\left((\bar{A}_{f}^{*})_{i}(A_{f})_{\perp}+(\bar{A}_{f}^{*})_{\perp}(A_{f})_{i}\right)\right]
-2xIm\left[\frac{p}{q}i\left((\bar{A}_{f}^{*})_{i}(A_{f})_{\perp}+(\bar{A}_{f}^{*})_{\perp}(A_{f})_{i}\right)\right]\nonumber\\
\bar{Y}^{V_{1}V_{2}}_{0\parallel}=&2yRe\left[\frac{p}{q}\left((\bar{A}_{f})_{0}^{*}(A_{f})_{\parallel}+(\bar{A}_{f})_{\parallel}^{*}(A_{f})_{0}\right)\right]
-2xIm\left[\frac{p}{q}\left((\bar{A}_{f})_{0}^{*}(A_{f})_{\parallel}+(\bar{A}_{f})_{\parallel}^{*}(A_{f})_{0}\right)\right]\nonumber\\
\bar{Y}^{\overline{V}_{1}\overline{V}_{2}}_{ii}=&2yRe\left[\frac{p}{q}(\bar{A}_{\bar f}^{*})_{i}(A_{\bar f})_{i}\right]
-2xIm\left[\frac{p}{q}(\bar{A}_{\bar f}^{*})_{i}(A_{\bar f})_{i}\right]\nonumber\\
\bar{Y}^{\overline{V}_{1}\overline{V}_{2}}_{\perp\perp}=&-2yRe\left[\frac{p}{q}(\bar{A}_{\bar f}^{*})_{\perp}(A_{\bar f})_{\perp}\right]
+2xIm\left[\frac{p}{q}(\bar{A}_{\bar f}^{*})_{\perp}(A_{\bar f})_{\perp}\right]\nonumber\\
\bar{Y}^{\overline{V}_{1}\overline{V}_{2}}_{i\perp}=&2yRe\left[\frac{p}{q}i\left((\bar{A}_{\bar f}^{*})_{i}(A_{\bar f})_{\perp}+(\bar{A}_{\bar f}^{*})_{\perp}(A_{\bar f})_{i}\right)\right]
-2xIm\left[\frac{p}{q}i\left((\bar{A}_{\bar f}^{*})_{i}(A_{\bar f})_{\perp}+(\bar{A}_{\bar f}^{*})_{\perp}(A_{\bar f})_{i}\right)\right]\nonumber\\
\bar{Y}^{\overline{V}_{1}\overline{V}_{2}}_{0\parallel}=&2yRe\left[\frac{p}{q}\left((\bar{A}_{\bar f})_{0}^{*}(A_{\bar f})_{\parallel}+(\bar{A}_{\bar f})_{\parallel}^{*}(A_{\bar f})_{0}\right)\right]
-2xIm\left[\frac{p}{q}\left((\bar{A}_{\bar f})_{0}^{*}(A_{\bar f})_{\parallel}+(\bar{A}_{\bar f})_{\parallel}^{*}(A_{\bar f})_{0}\right)\right]
\end{align}
where $\frac{q}{p}=re^{i(\alpha_{SM}+\alpha_{NP})} \footnote{Within the SM, CP violation in the charm sector is known to be very small ($\mathcal O(\lambda^5)$ in the Wolfenstein parameterization of the CKM matrix). However, no NP has yet been found in the charm sector, and hence, any NP effect, if present, has to be very tiny too. It is therefore not feasible to neglect the tiny CP violating phase in charm mixing when one tries to probe NP in the charm sector. Hence we parameterize the CP violating mixing phase in charm according to the SM ($\alpha_{SM}$) and the NP ($\alpha_{NP}$) counterparts.}$.
These 48 observables can be written in terms of 28 independent parameters: $r$,
$\alpha_{SM}, \alpha_{NP}$, $\phi$, $\chi$, three each of  $a_{\lambda}$'s, $b_{\lambda}$'s, $c_{\lambda}$'s, $d_{\lambda}$'s, and eleven strong phase differences $\delta^{ab}_{\lambda}=\delta^{a}_{\lambda}-\delta^{b}_{\lambda}$,
$\delta^{cd}_{\lambda}=\delta^{c}_{\lambda}-\delta^{d}_{\lambda}$, $\delta^{ac}_{\lambda}=\delta^{a}_{\lambda}-\delta^{c}_{\lambda}$,
$\Delta^{a}_{\parallel}=\delta^{a}_{\perp}-\delta^{a}_{\parallel}$. The expressions of the observables in terms of the theoretical parameters can be found in appendix~\ref{obs}.

\section{The $\textbf{a}$, $\textbf{b}$, $\textbf{c}$ amplitudes\label{pol}}
In this section, we briefly discuss the relation of the polarization basis to the $a$, $b$ and $c$ amplitudes used in the most general covariant expression for a $D\rightarrow V_1V_2$ decay. We closely follow~\cite{Valencia} in the discussions of this section.

Consider the decay $D^{0}\to V_{1}V_{2}$. Angular momentum conservation ($\vec{J}_{D}=\vec{L}_{V_{1}V_{2}}+\vec{S}_{V_{1}V_{2}}$), dictates that $\vec{L}_{V_{1}V_{2}}$ can be 0, 1, 2.
Thus, one obtains three independent amplitudes corresponding to the three different $\vec{L}_{V_{1}V_{2}}$ values. The most general covariant amplitude for a $D^{0}\to V_{1}V_{2}$ decay can be written as~\cite{Valencia,Lipkin}
\begin{align}\label{covariant amplitude}
 M_{\lambda_{1}\lambda_{2}}=&<V_{1}(\lambda_{1})V_{2}(\lambda_{2})|H_{wk}|D>\nonumber\\
=&\epsilon_{1\mu}^{*}\epsilon_{2\nu}^{*}\bigg[ag^{\mu\nu}+\frac{b}{m_{1}m_{2}}p_{2}^{\mu}p_{1}^{\nu}+i\frac{c}{m_{1}m_{2}}\epsilon^{\mu\nu\alpha\beta}p_{1\alpha}p_{2\beta}\bigg],
\end{align}
Here $\epsilon_{1}$, $\epsilon_{2}$ represent the polarization vectors and $m_{1}$, $m_{2}$ the masses of the vector mesons $V_{1}$ and $V_{2}$ respectively. The invariant amplitudes  $a$, $b$, $c$ each carry the dimension of energy.
The corresponding decay rate in terms of the $a$, $b$ and $c$ amplitudes is given by
\begin{equation}\label{gamma_cov}
 \Gamma\left(D\to V_{1}V_{2}\right)=\frac{\left|\bf k\right|}{8\pi m_{D}^{2}}\left(2|a|^{2}+|xa+(x^{2}-1)b|^{2}+2(x^{2}-1)|c|^{2}\right],
\end{equation}
where $\left|\bf k\right|$ is the decay momentum and $x=\frac{p_{1}.p_{2}}{m_{1}m_{2}}=\frac{m_{D}^{2}-m_{1}^{2}-m_{2}^{2}}{2m_{1}m_{2}}$.

The $D^{0}\to V_{1}V_{2}$ amplitude can also be written in the linear polarization basis as,
\begin{align}\label{linear polarisation amplitude}
 A\left(D^{0}\to V_{1}V_{2}\right)=A_{0}\epsilon_{1}^{*L}\epsilon_{2}^{*L}-A_{\parallel}\vec{\epsilon}_{1}^{*T}.\vec{\epsilon}_{2}^{*T}/\sqrt{2}
 -iA_{\perp}\vec{\epsilon}_{1}^{*}\times\vec{\epsilon}_{2}^{*}.\hat{p}_{2}/\sqrt{2},
\end{align}
where $\epsilon_{i}^{*L}=\vec{\epsilon}_{i}^{*}.\hat{p}_{2}$,  $\vec{\epsilon}_{i}^{*T}=\vec{\epsilon}_{i}^{*}-\epsilon_{i}^{*L}\hat{p}_{2}$, $\hat{p}_{2}$ is a unit vector in the direction of the momentum of the meson $V_{2}$ in the $V_{1}$
rest frame. Comparing eqn.(\ref{linear polarisation amplitude}) to eqn.(\ref{amplitude}), it can easily be verified that $g_{0}=\epsilon_{1}^{*L}\epsilon_{2}^{*L}$, $g_{\parallel}=-\frac{1}{\sqrt{2}}\epsilon_{1}^{*T}.\epsilon_{2}^{*T}$ and 
$g_{\perp}=-\frac{1}{\sqrt{2}}\epsilon_{1}^{*}\times\epsilon_{2}^{*}.\hat{p}_{2}$. The corresponding decay width in the linear polarization basis is given by
\begin{equation}\label{gamma_pol}
 \Gamma\left(D^{0}\to V_{1}V_{2}\right)=\frac{\left|\bf k\right|}{8\pi m_{D}^{2}}\left(|A_{0}|^{2}+|A_{\parallel}|^{2}+|A_{\perp}|^{2}\right).
\end{equation}
From eqns.(\ref{gamma_cov}) and (\ref{gamma_pol}), it is evident that the two bases are related as
\begin{eqnarray}
 A_{\parallel}=\sqrt{2}a,\,\,\,A_{0}=-ax-b(x^{2}-1),\,\,\,A_{\perp}=\sqrt{2(x^{2}-1)}c.
\end{eqnarray}

\section{Results and discussions\label{results}}
 A priori, the dependence of the observables on the theoretical parameters may not be trivial as is evident from appendix~\ref{obs}. However, a careful and systematic study of these observables enables one to propose multiple ways for the identification and extraction of new physics as will be shown in this section. In what follows, $\lambda=\{0,\parallel,\perp\}$ and $i=\{0,\parallel\}$.
 
 Let us look at the SM case first. For this case the NP decay amplitudes $b_{\lambda}$, $d_{\lambda}$, the NP strong phases $\delta^{b}_{\lambda}$, $\delta^{d}_{\lambda}$ and the NP weak phases $\phi$ and $\chi$ and $\alpha_{NP}$ are all equal to 0.  With these values, we get the following relations between the observables
  \begin{align}
  &X^{V_1V_2}_{\lambda\lambda}=\bar{X}^{\bar{V_1}\bar{V_2}}_{\lambda\lambda},\;
  X^{\bar{V_1}\bar{V_2}}_{\lambda\lambda}=\bar{X}^{V_1V_2}_{\lambda\lambda},\;
  X^{V_1V_2}_{0\parallel}=\bar{X}^{\bar{V_1}\bar{V_2}}_{0\parallel},\label{SM_1}\\
  &X^{\bar{V_1}\bar{V_2}}_{0\parallel}=\bar{X}^{V_1V_2}_{0\parallel},\;
  X^{V_1V_2}_{i\perp}=-\bar{X}^{\bar{V_1}\bar{V_2}}_{i\perp},\;
  X^{\bar{V_1}\bar{V_2}}_{i\perp}=-\bar{X}^{V_1V_2}_{i\perp}\label{SM_2}.
 \end{align}
 It is possible to construct linear combinations of the $Y$'s which depend directly on the phases $\alpha_{SM}$ and the decay phase difference $\delta_{\lambda}^{ac}$ as follows:
 \begin{align}
  &1+\frac{Y_{\lambda \lambda }^{\bar{V_1} \bar{V_2}}+Y_{\lambda \lambda }^{V_1
   V_2}}{r^2 \bar{Y}_{\lambda \lambda }^{\bar{V_1} \bar{V_2}}+r^2 \bar{Y}_{\lambda
   \lambda }^{V_1 V_2}}=\frac{2y\cos\alpha_{SM}}{y\cos\alpha_{SM}+x\sin\alpha_{SM}}\label{SM_3}\\
  &\frac{r^2 \bar{Y}_{\lambda \lambda }^{\bar{V_1} \bar{V_2}}+r^2 \bar{Y}_{\lambda
   \lambda }^{V_1 V_2}+Y_{\lambda \lambda }^{\bar{V_1} \bar{V_2}}+Y_{\lambda
   \lambda }^{V_1 V_2}}{r^2 \bar{Y}_{\lambda \lambda }^{\bar{V_1} \bar{V_2}}-r^2
   \bar{Y}_{\lambda \lambda }^{V_1 V_2}-Y_{\lambda \lambda }^{\bar{V_1}
   \bar{V_2}}+Y_{\lambda \lambda }^{V_1 V_2}}=\frac{x}{y}\tan\delta_{\lambda}^{ac}\label{SM_4}
 \end{align}
 Eqn.(\ref{SM_3}) is of particular interest. The right hand side of the equation is known to appreciable accuracy for the SM, and so is $r$ on the left hand side. The observables $Y$ can be measured experimentally. Hence this relation can be used as a smoking gun signal for detecting the presence of NP.
 
 One can go further and identify whether the NP manifests itself in pure decays or as a CP violating effect in mixing or both. For instance, suppose there is no NP in pure decays but some signal for CP violation has been observed. In that case, the formalism dictates that $b_{\lambda}=d_{\lambda}=\delta^{b}_{\lambda}=\delta^{d}_{\lambda}=\phi=\chi=0$, but $\alpha_{NP}\neq0$. 
 
 This case is very similar to the SM case discussed above. In particular the relations (\ref{SM_1}), (\ref{SM_2}) and (\ref{SM_4}) hold exactly in the same way. This is because the $X$'s do not depend on the (small) NP phase $\alpha_{NP}$ at all, and in the other case the $\alpha_{NP}$ dependence gets canceled in the left hand side of eqn.(\ref{SM_4}). However eqn.(\ref{SM_3}) is modified to 
 \begin{equation}
1+\frac{Y_{\lambda \lambda }^{\bar{V_1} \bar{V_2}}+Y_{\lambda \lambda }^{V_1
   V_2}}{r^2 \bar{Y}_{\lambda \lambda }^{\bar{V_1} \bar{V_2}}+r^2 \bar{Y}_{\lambda
   \lambda }^{V_1 V_2}}=\frac{2y}{y+x\tan(\alpha_{SM}+\alpha_{NP})}\label{no_decay_1}\\
 \end{equation}
 Eqn.(\ref{no_decay_1}) along with relations of the type
 \begin{equation}
  Y^{V_1V_2}_{\lambda\lambda}+r^2\bar{Y}^{\bar{V_1}\bar{V_2}}_{\lambda\lambda}+Y^{\bar{V_1}\bar{V_2}}_{\lambda\lambda}+r^2\bar{Y}^{V_1V_2}_{\lambda\lambda}=4y\frac{\cos\alpha_{SM}+\alpha_{NP}}{y\cos\alpha_{SM}+\alpha_{NP}-x\sin\alpha_{SM}+\alpha_{NP}}\frac{Y^{V_1V_2}_{\lambda\lambda}+Y^{\bar{V_1}\bar{V_2}}_{\lambda\lambda}}{\sqrt{X^{V_1V_2}_{\lambda\lambda}X^{\bar{V_1}\bar{V_2}}_{\lambda\lambda}}}.
 \end{equation}
 can be simultaneously solved for the determination of $r$ and $\alpha_{NP}$ in this case.

 We next look into the case where NP is manifested only in pure decays. In this case, $\alpha_{NP}=0$, but $b_{\lambda}$, $d_{\lambda}$, $\delta^{b}_{\lambda}$, $\delta^{d}_{\lambda}$, $\phi$, $\chi$ $\neq0$. For these values of the parameters, we have the following relations among the observables $X$ and $Y$:
 \begin{equation}
  \frac{x^2({Y^{V_1V_2}_{\lambda\lambda}}+{\bar{Y}^{V_1V_2}_{\lambda\lambda}})^2+y^2({Y^{V_1V_2}_{\lambda\lambda}}-{\bar{Y}^{V_1V_2}_{\lambda\lambda}})^2}{x^2({Y^{\bar{V_1}\bar{V_2}}_{\lambda\lambda}}+{\bar{Y}^{\bar{V_1}\bar{V_2}}_{\lambda\lambda}})^2+y^2({Y^{\bar{V_1}\bar{V_2}}_{\lambda\lambda}}-{\bar{Y}^{\bar{V_1}\bar{V_2}}_{\lambda\lambda}})^2}=\frac{\bar{X}^{V_1V_2}_{\lambda\lambda}{X^{V_1V_2}_{\lambda\lambda}}}{{\bar{X}^{\bar{V_1}\bar{V_2}}_{\lambda\lambda}}{X^{\bar{V_1}\bar{V_2}}_{\lambda\lambda}}}.\label{no_mixing_1}
 \end{equation}
 It would be appropriate here to point out relations of the type
 \begin{eqnarray}
  &{X^{V_1V_2}_{\lambda\lambda}}-{\bar{X}^{\bar{V_1}\bar{V_2}}_{\lambda\lambda}}=8\sin\phi\sin\delta^{ab}_{\lambda}a_{\lambda}b_{\lambda},\label{no_mixing_2}\\
  &{X^{\bar{V_1}\bar{V_2}}_{\lambda\lambda}}-{\bar{X}^{V_1V_2}_{\lambda\lambda}}=8\sin\chi\sin\delta^{cd}_{\lambda}c_{\lambda}d_{\lambda}\label{no_mixing_3}.
 \end{eqnarray}
 Note that a non zero value of either of the above linear combinations is again a smoking gun signal for NP, since that will mean that the NP weak phases ($\phi$, $\chi$) and the NP weak decay amplitudes ($b_{\lambda}$, $d_{\lambda}$) $\neq0$.

 Eqns.(\ref{no_mixing_1}), (\ref{no_mixing_2}) and (\ref{no_mixing_3}) hold in the most general case also were NP is present both in pure decays and mixing. The extraction of $\mid\frac{q}{p}\mid$ in the most general scenario can be obtained from relations of the type:
 \begin{equation}
   x^2({Y^{V_1V_2}_{\lambda\lambda}}+{r^2\bar{Y}^{V_1V_2}_{\lambda\lambda}})^2+y^2({Y^{V_1V_2}_{\lambda\lambda}}-{r^2\bar{Y}^{V_1V_2}_{\lambda\lambda}})^2=4r^2x^2y^2{\bar{X}^{V_1V_2}_{\lambda\lambda}}{X^{V_1V_2}_{\lambda\lambda}}.
  \end{equation}

 Let us now discuss if these observables can be used for the extraction of all the parameters. It is appropriate at this point to look into the case of final states in $D^0$ decays that are CP eigenstates. These include modes like $D^0\rightarrow\phi\omega$, $D^0\rightarrow\rho^0\phi$, $D^0\rightarrow\rho^0\omega$, $D^0\rightarrow\rho\rho$, $D^0\rightarrow\omega\omega$ etc. For these modes, $O^{V_1V_2}=O^{\bar{V_1}\bar{V_2}}$ where $O=\{X, \bar{X}, Y, \bar{Y}\}$ for all combinations of the polarization indices. Therefore the number of observables reduces from 48 to 24. However there are 28 parameters. Discarding $\alpha_{SM}$ (which is precisely known already) and $r$ (since the NP effect, if  present, must be very small, it is a reasonable approximation to incorporate the complete effect of NP for the ratio $q/p$ into the phase $\alpha_{NP}$ without changing $r$, which, then, is again precisely known) still leaves 26 parameters to be fitted to 24 observables. Hence, the numerical estimates for all the parameters cannot be obtained for final states that are CP eigenstates without some approximations to reduce the number of parameters. However, relations like (\ref{no_mixing_2}) (eqn.(\ref{no_mixing_2}) and eqn.(\ref{no_mixing_3}) are same in this case) where linear combinations of experimentally measured observables vanish in the absence of new physics can still be devised as tests for the detection of NP effects. For a complete estimation of all the parameters, one has to look at final states which are non CP eigenstates, i.e. modes involving $K^{*0}$ as one of the final states.
 
 At present, there is serious dearth of data in the $D\rightarrow VV$ sector. The most well measured mode is $D^0\rightarrow\bar{K^{*0}}\rho^0$ where there are estimates for the total branching fraction, the total transverse wave, S-wave, longitudinal S-wave, P-wave and D-wave branching fractions~\cite{pdg}. The P-wave and longitudinal branching fraction are limits. This gives one six branching fractions which can be written in terms of the theoretical parameters. With the corresponding measurements of $D^0\rightarrow K^{*0}\rho^0$, $\bar{D^0}\rightarrow K^{*0}\rho^0$ and $\bar{D^0}\rightarrow \bar{K^{*0}}\rho^0$ (which are well within the present experimental reach), one has 24 branching fractions written in terms of 27 parameters (discarding $\alpha_{SM}$). It is hence not possible to obtain numerical estimates for all the 27 theoretical parameters with the present data set. However with the measurement of another observable in the future, (say the total longitudinal wave), it is possible to extract numerical
 values for all the theoretical parameters from a fit to data. 
 
 With the present data set, one can find numerical estimates for the absolute values of the $A_0$, $A_{\parallel}$ and the $A_{\perp}$ amplitudes (and the corresponding observables $X_{\lambda\lambda}$ from the first relation in (\ref{pol_obs})) and the interference phase between $A_0$ and $A_{\perp}$ for the $D^0\rightarrow\bar{K^{*0}}\rho^0$ mode. We have extracted the amplitudes and the interference phase (which we denote by $\theta$) via a numerical fit using the package HEPfit~\cite{hepfit}. The values of the amplitudes and the phase are displayed in table~\ref{tab1}.
 \begin{table}[ht]
\captionsetup{justification=raggedright,
singlelinecheck=false
}
\centering
\begin{tabular}{|c|c|}\hline
~Amplitude~&~Values~\\
\hline\hline 
$\mid A_0\mid$        &$(4.758\pm3.052)\times10^{-7}$\\\hline
$\mid A_\parallel\mid$&$(8.025\pm4.724e)\times10^{-7}$\\\hline
$\mid A_\perp\mid$    &$(2.0671\pm0.19355)\times10^{-6}$\\\hline
$\theta$              &$(0.00007\pm0.33)^{\circ}$\\\hline
\hline
\end{tabular}
\caption{The absolute values for the longitudinal ($A_0$), parallel transverse ($A_\parallel$) and perpendicular transverse ($A_\perp$) amplitudes and the interference phase between the longitudinal and the perpendicular transverse amplitudes ($\theta$) for $D\rightarrow\overline{K^{*0}}\rho^0$. The values have been extracted from a fit to data. The data is taken from~\cite{pdg}.}
\label{tab1}
\end{table}
\begin{table}[ht]
\captionsetup{justification=raggedright,
singlelinecheck=false
}
\centering
\begin{tabular}{|c|c|c|}\hline
~Branching ratios~&~Our values~&~Experimental values~\\
\hline\hline 
S-wave     &$(2.79\pm0.51)\%$&$(3.00\pm0.60)\%$\\\hline
D-wave     &$(2.21\pm0.46)\%$&$(2.10\pm0.60)\%$\\\hline
Transverse &$(1.55\pm0.26)\%$&$(1.70\pm0.60)\%$\\\hline
Total      &$(1.65\pm0.24)\%$&$(1.57\pm0.35)\%$\\\hline
\hline
\end{tabular}
\caption{The total branching ratio and the branching ratios for the various partial waves for the $D\rightarrow\overline{K^{*0}}\rho^0$ mode. The values due to our fit are displayed in the second column and the experimental values are displayed in the last column. All the experimental values are taken from~\cite{pdg}.}
\label{tab2}
\end{table}
 
 The corresponding values thus obtained for the branching ratios are given in table~\ref{tab2}. These values are consistent with experiment within errors. 

\section{Summary and conclusions\label{summary}}
We present a model independent formalism for the detection of NP in $D^0\rightarrow VV$ decays, which involves a time-dependent analysis of the neutral $D$ mesons. This formalism has been previously used to analyze the $B\rightarrow VV$ decays. However the form of the observables change due to the different parametrization of the time dependent $D^0$ decays. We show that it is possible to construct combinations of these observables which are smoking gun signals for NP. We discuss the relations between the observables in the different cases of the SM, NP in decay, NP in mixing and the most general case where NP manifests itself both in mixing and decay. We also extract the absolute values for the longitudinal, the transverse parallel, the transverse perpendicular amplitudes and the interference phase between the longitudinal and the transverse perpendicular amplitudes for the $D^0\rightarrow\overline{K^{*0}}\rho^0$ mode. We find that the interference phase is consistent with zero.
 
\acknowledgements 

We thank Nita Sinha and Rahul Sinha for fruitful discussions regarding this work.
 \appendix
 
\section{Observables extracted from the angular analysis}\label{obs}
 We list all the 48 observables in terms of the 28 parameters for the most general case. All the relations discussed in section~\ref{results} follow from these. In the following, $\lambda={0, \perp, \parallel}$ and $i=0, \parallel$.
\begin{align*}
X_{\lambda\lambda}^{\text{V}_1\text{V}_2}= &2 \left(2 a_\lambda b_\lambda \cos \left(\phi -\delta _\lambda^{\text{ab}}\right)+a_\lambda^2+b_\lambda^2\right)\\
X_{i\perp}^{\text{V}_1\text{V}_2}= &4 \left(a_i \left(a_\perp \left(-\sin \left(\Delta _i^a\right)\right)-b_\perp \sin \left(-\Delta _i^a+\delta _\perp^{\text{ab}}+\phi \right)\right)
                                        +b_i \left(a_\perp \sin \left(-\Delta _i^a-\delta _i^{\text{ab}}+\phi \right)\right.\right.\\
                                        &\left.\left.-b_\perp \sin \left(\Delta _i^a+\delta _i^{\text{ab}}-\delta _\perp^{\text{ab}}\right)\right)\right)\\
\end{align*}
\begin{align*}
X_{0\parallel}^{\text{V}_1\text{V}_2}= &4 \left(a_0 \left(b_\parallel \cos \left(\Delta _0^a-\Delta _\parallel^a-\delta _\parallel^{\text{ab}}+\phi \right)+a_\parallel \cos \left(\Delta _0^a-\Delta _\parallel^a\right)\right)+b_0 \left(b_\parallel \cos \left(\Delta _0^a-\Delta _\parallel^a+\delta _0^{\text{ab}}-\delta _\parallel^{\text{ab}}\right)\right.\right.\\
&\left.\left.+a_\parallel \cos \left(-\Delta _0^a+\Delta _\parallel^a-\delta _0^{\text{ab}}+\phi \right)\right)\right)\\                                        
X_{\lambda\lambda}^{\bar{\text{V}_1}\bar{\text{V}_2} }= &2 \left(2 c_\lambda d_\lambda \cos \left(\gamma -\delta _\lambda^{\text{cd}}\right)+c_\lambda^2+d_\lambda^2\right)\\
X_{i\perp}^{\bar{\text{V}_1}\bar{\text{V}_2} }= &4 \left(c_i \left(c_\perp \left(-\sin \left(\Delta _i^a+\delta _i^{\text{ac}}-\delta _\perp^{\text{ac}}\right)\right)-d_\perp \sin \left(\Delta _i^a+\delta _i^{\text{ac}}-\delta _\perp^{\text{ac}}+\gamma -\delta _\perp^{\text{cd}}\right)\right)\right.\\
                                     &\left.+d_i \left(c_\perp \sin \left(-\Delta _i^a-\delta _i^{\text{ac}}+\delta _\perp^{\text{ac}}+\gamma -\delta _\lambda^{\text{cd}}\right)
                                     -d_\perp \sin \left(\Delta _i^a+\delta _i^{\text{ac}}-\delta _\perp^{\text{ac}}+\delta _\lambda^{\text{cd}}-\delta _\perp^{\text{cd}}\right)\right)\right)\\
X_{0\parallel}^{\bar{\text{V}_1}\bar{\text{V}_2}}= &4 \left(c_0 \left(c_\parallel \cos \left(\Delta _0^a-\Delta _\parallel^a+\delta _0^{\text{ac}}-\delta _\parallel^{\text{ac}}\right)+d_\parallel \cos \left(\Delta _0^a-\Delta _\parallel^a+\delta _0^{\text{ac}}-\delta _\parallel^{\text{ac}}+\gamma -\delta _\parallel^{\text{cd}}\right)\right)+d_0 \left(c_\parallel\right.\right.\\
&\left.\left.\cos \left(-\Delta _0^a+\Delta _\parallel^a-\delta _0^{\text{ac}}+\delta _\parallel^{\text{ac}}+\gamma -\delta _0^{\text{cd}}\right)+d_\parallel \cos \left(\Delta _0^a-\Delta _\parallel^a+\delta _0^{\text{ac}}-\delta _\parallel^{\text{ac}}+\delta _0^{\text{cd}}-\delta _\parallel^{\text{cd}}\right)\right)\right)\\
\bar{X}_{\lambda\lambda}^{\text{V}_1\text{V}_2}= &2 \left(2 c_\lambda d_\lambda \cos \left(\gamma +\delta _\lambda^{\text{cd}}\right)+c_\lambda^2+d_\lambda^2\right)\\
\bar{X}_{i\perp}^{\text{V}_1\text{V}_2}= &-4 \left(c_i \left(d_\perp \sin \left(-\Delta _i^a-\delta _i^{\text{ac}}+\delta _\perp^{\text{ac}}+\gamma +\delta _\perp^{\text{cd}}\right)-c_\perp \sin \left(\Delta _i^a+\delta _i^{\text{ac}}-\delta _\perp^{\text{ac}}\right)\right)\right.\\
                                     &\left.+d_i \left(c_\perp \left(-\sin \left(\Delta _i^a+\delta _i^{\text{ac}}-\delta _\perp^{\text{ac}}+\gamma +\delta _\lambda^{\text{cd}}\right)\right)
                                     -d_\perp \sin \left(\Delta _i^a+\delta _i^{\text{ac}}-\delta _\perp^{\text{ac}}+\delta _\lambda^{\text{cd}}-\delta _\perp^{\text{cd}}\right)\right)\right)\\
\bar{X}_{0\parallel}^{\text{V}_1\text{V}_2}= &4 \left(d_0 \left(c_\parallel \cos \left(\Delta _0^a-\Delta _\parallel^a+\delta _0^{\text{ac}}-\delta _\parallel^{\text{ac}}+\gamma +\delta _0^{\text{cd}}\right)+d_\parallel \cos \left(\Delta _0^a-\Delta _\parallel^a+\delta _0^{\text{ac}}-\delta _\parallel^{\text{ac}}+\delta _0^{\text{cd}}-\delta _\parallel^{\text{cd}}\right)\right)\right.\\
&\left.+c_0 \left(c_\parallel \cos \left(\Delta _0^a-\Delta _\parallel^a+\delta _0^{\text{ac}}-\delta _\parallel^{\text{ac}}\right)+d_\parallel \cos \left(-\Delta _0^a+\Delta _\parallel^a-\delta _0^{\text{ac}}+\delta _\parallel^{\text{ac}}+\gamma +\delta _\parallel^{\text{cd}}\right)\right)\right)\\
\bar{X}_{\lambda\lambda}^{\bar{\text{V}_1}\bar{\text{V}_2} }= &2 \left(2 a_\lambda b_\lambda \cos \left(\delta_\lambda^{\text{ab}}+\phi \right)+a_\lambda^2+b_\lambda^2\right)\\
\bar{X}_{i\perp}^{\bar{\text{V}_1}\bar{\text{V}_2} }= &-4 \left(a_i \left(b_\perp \sin \left(\Delta_i^a-\delta _\perp^{\text{ab}}+\phi \right)-a_\perp \sin \left(\Delta_i^a\right)\right)+b_i\left(a_\perp \left(-\sin \left(\Delta _i^a+\delta _i^{\text{ab}}+\phi \right)\right)\right.\right.\\
                                     &\left.\left.-b_\perp \sin \left(\Delta _i^a+\delta _i^{\text{ab}}-\delta _\perp^{\text{ab}}\right)\right)\right)\\
\bar{X}_{0\parallel}^{\bar{\text{V}_1}\bar{\text{V}_2}}= &4 \left(b_0 \left(b_\parallel \cos \left(\Delta _0^a-\Delta _\parallel^a+\delta _0^{\text{ab}}-\delta _\parallel^{\text{ab}}\right)+a_\parallel \cos \left(\Delta _0^a-\Delta _\parallel^a+\delta _0^{\text{ab}}+\phi \right)\right)+a_0 \left(b_\parallel \cos \left(-\Delta _0^a+\Delta _\parallel^a\right.\right.\right.\\
&\left.\left.\left.+\delta _\parallel^{\text{ab}}+\phi \right)+a_\parallel \cos \left(\Delta _0^a-\Delta _\parallel^a\right)\right)\right)\\
Y_{ii}^{\text{V}_1\text{V}_2}= &2 r \left(a_i \left(d_i \left(y \cos \left(\alpha_{SM}+\alpha_{NP} -\delta _i^{\text{ac}}-\gamma -\delta _\lambda^{\text{cd}}\right)-x \sin \left(\alpha_{SM}+\alpha_{NP} -\delta _i^{\text{ac}}-\gamma -\delta _\lambda^{\text{cd}}\right)\right)+c_i \left(y \cos\right.\right.\right.\\ &\left.\left.\left.\left(\alpha_{SM}+\alpha_{NP} -\delta _i^{\text{ac}}\right)
-x \sin \left(\alpha_{SM}+\alpha_{NP} -\delta _i^{\text{ac}}\right)\right)\right)+b_i \left(d_i \left(y \cos \left(\alpha_{SM}+\alpha_{NP} +\delta _i^{\text{ab}}-\delta _i^{\text{ac}}-\gamma \right.\right.\right.\right.\\
&\left.\left.\left.\left.-\delta _\lambda^{\text{cd}}-\phi \right)-x \sin \left(\alpha_{SM}+\alpha_{NP} +\delta _i^{\text{ab}}-\delta _i^{\text{ac}}-\gamma -\delta _\lambda^{\text{cd}}-\phi \right)\right)+c_i \left(y \cos \left(\alpha_{SM}+\alpha_{NP} +\delta _i^{\text{ab}}-\delta _i^{\text{ac}}\right.\right.\right.\right.\\
&\left.\left.\left.\left.-\phi \right)-x \sin \left(\alpha_{SM}+\alpha_{NP} +\delta _i^{\text{ab}}-\delta _i^{\text{ac}}-\phi \right)\right)\right)\right)\\
Y_{\perp\perp}^{\text{V}_1\text{V}_2}= &-2 r \left(a_\perp \left(d_\perp \left(y \cos \left(\alpha_{SM}+\alpha_{NP} -\delta _\perp^{\text{ac}}-\gamma -\delta _\perp^{\text{cd}}\right)-x \sin \left(\alpha_{SM}+\alpha_{NP} -\delta _\perp^{\text{ac}}-\gamma -\delta _\perp^{\text{cd}}\right)\right)+c_\perp \right.\right.\\
&\left.\left.\left(y \cos \left(\alpha_{SM}+\alpha_{NP} -\delta _\perp^{\text{ac}}\right)-x \sin \left(\alpha_{SM}+\alpha_{NP} -\delta _\perp^{\text{ac}}\right)\right)\right)+b_\perp \left(d_\perp \left(y \cos \left(\alpha_{SM}+\alpha_{NP} +\delta _\perp^{\text{ab}}-\delta _\perp^{\text{ac}}\right.\right.\right.\right.\\
&\left.\left.\left.\left.-\gamma -\delta _\perp^{\text{cd}}-\phi \right)-x \sin \left(\alpha_{SM}+\alpha_{NP} +\delta _\perp^{\text{ab}}-\delta _\perp^{\text{ac}}-\gamma -\delta _\perp^{\text{cd}}-\phi \right)\right)+c_\perp \left(y \cos \left(\alpha_{SM}+\alpha_{NP} +\delta _\perp^{\text{ab}}\right.\right.\right.\right.\\
&\left.\left.\left.\left.-\delta _\perp^{\text{ac}}-\phi \right)-x \sin \left(\alpha_{SM}+\alpha_{NP} +\delta _\perp^{\text{ab}}-\delta _\perp^{\text{ac}}-\phi \right)\right)\right)\right)\\
Y_{i\perp}^{\text{V}_1\text{V}_2}= &2 r \left(b_\perp \left(d_i \left(x \cos \left(-\Delta _i^a+\alpha_{SM}+\alpha_{NP} +\delta _\perp^{\text{ab}}-\delta _i^{\text{ac}}-\gamma -\delta _\lambda^{\text{cd}}-\phi \right)+y \sin \left(-\Delta _i^a+\alpha_{SM}+\alpha_{NP} \right.\right.\right.\right.\\
&\left.\left.\left.\left.+\delta _\perp^{\text{ab}}-\delta _i^{\text{ac}}-\gamma -\delta _\lambda^{\text{cd}}-\phi \right)\right)+c_i \left(x \cos \left(-\Delta _i^a+\alpha_{SM}+\alpha_{NP} +\delta _\perp^{\text{ab}}-\delta _i^{\text{ac}}-\phi \right)+y \sin \left(-\Delta _i^a+ \right.\right.\right.\right.\\
\end{align*}
\begin{align*}
&\left.\left.\left.\left.\alpha_{SM}+\alpha_{NP} +\delta _\perp^{\text{ab}}-\delta _i^{\text{ac}}-\phi \right)\right)\right)+d_\perp \left(b_i \left(x \cos \left(\Delta _i^a+\alpha_{SM}+\alpha_{NP} -\delta _i^{\text{ab}}-\delta _\perp^{\text{ac}}-\gamma -\delta _\perp^{\text{cd}}-\phi \right)\right.\right.\right.\\
&\left.\left.\left.+y \sin \left(\Delta _i^a+\alpha_{SM}+\alpha_{NP} -\delta _i^{\text{ab}}-\delta _\perp^{\text{ac}}-\gamma -\delta _\perp^{\text{cd}}-\phi \right)\right)+a_i \left(x \cos \left(\Delta _i^a+\alpha_{SM}+\alpha_{NP} -\delta _\perp^{\text{ac}}-\gamma\right.\right.\right.\right.
&\left.\left.\left.\left.-\delta _\perp^{\text{cd}}\right)+y \sin \left(\Delta _i^a+\alpha_{SM}+\alpha_{NP} -\delta _\perp^{\text{ac}}-\gamma -\delta _\perp^{\text{cd}}\right)\right)\right)\right.\\
&\left.+c_\perp \left(b_i \left(x \cos \left(\Delta _i^a+\alpha_{SM}+\alpha_{NP} +\delta _i^{\text{ab}}-\delta _\perp^{\text{ac}}-\phi \right)+y \sin \left(\Delta _i^a+\alpha_{SM}+\alpha_{NP} +\delta _i^{\text{ab}}-\delta _\perp^{\text{ac}}-\phi \right)\right)\right.\right.\\
&\left.\left.+a_i \left(x \cos \left(\Delta _i^a+\alpha_{SM}+\alpha_{NP} -\delta _\perp^{\text{ac}}\right)+y \sin \left(\Delta _i^a+\alpha_{SM}+\alpha_{NP} -\delta _\perp^{\text{ac}}\right)\right)\right)+a_\perp \left(d_i \left(x \cos \left(-\Delta _i^a+\right.\right.\right.\right.\\
&\left.\left.\left.\left.\alpha_{SM}+\alpha_{NP} -\delta _i^{\text{ac}}-\gamma -\delta _\lambda^{\text{cd}}\right)
y \sin \left(-\Delta _i^a+\alpha_{SM}+\alpha_{NP} -\delta _i^{\text{ac}}-\gamma -\delta _\lambda^{\text{cd}}\right)\right)+c_i \left(x \cos \left(-\Delta _i^a+\right.\right.\right.\right.\\
&\left.\left.\left.\left.+\alpha_{SM}+\alpha_{NP} -\delta _i^{\text{ac}}\right)+y \sin \left(-\Delta _i^a+\alpha_{SM}+\alpha_{NP} -\delta _i^{\text{ac}}\right)\right)\right)\right)\\
 Y_{0\parallel}^{\text{V}_1\text{V}_2}= &2 r \left(d_\parallel \left(a_0 \left(y \cos \left(\alpha_{SM}+\alpha_{NP} -\gamma -\text{$\delta $a}_0+\text{$\delta $d}_\parallel\right)-x \sin \left(\alpha_{SM}+\alpha_{NP} -\gamma -\text{$\delta $a}_0+\text{$\delta $d}_\parallel\right)\right)+b_0 \left(y \right.\right.\right.\\
 &\left.\left.\left.\cos \left(\alpha_{SM}+\alpha_{NP} -\gamma -\text{$\delta $b}_0+\text{$\delta $d}_\parallel-\phi \right)-x \sin \left(\alpha_{SM}+\alpha_{NP} -\gamma -\text{$\delta $b}_0+\text{$\delta $d}_\parallel-\phi \right)\right)\right)+c_\parallel\left(a_0 \left(y \right.\right.\right.\\
 &\left.\left.\left.\cos \left(\alpha_{SM}+\alpha_{NP} -\text{$\delta $a}_0+\text{$\delta $c}_\parallel\right)-x \sin \left(\alpha_{SM}+\alpha_{NP}
-\text{$\delta $a}_0+\text{$\delta $c}_\parallel\right)\right)+b_0 \left(y \cos \left(\alpha_{SM}+\alpha_{NP} -\text{$\delta $b}_0\right.\right.\right.\right.\\
&\left.\left.\left.\left.+\text{$\delta $c}_\parallel-\phi \right)-x \sin \left(\alpha_{SM}+\alpha_{NP} -\text{$\delta $b}_0+\text{$\delta $c}_\parallel-\phi \right)\right)\right)+a_\parallel \left(d_0 \left(y \cos 
 \left(\alpha_{SM}+\alpha_{NP} -\gamma -\text{$\delta $a}_\parallel+\right.\right.\right.\right.\\
 &\left.\left.\left.\left.\text{$\delta $d}_0\right)-x \sin \left(\alpha_{SM}+\alpha_{NP} -\gamma -\text{$\delta $a}_\parallel+\text{$\delta $d}_0\right)\right)+c_0 \left(y \cos \left(\alpha_{SM}+\alpha_{NP} -\text{$\delta $a}_\parallel+\text{$\delta $c}_0\right)-x \sin \left(\alpha_{SM}\right.\right.\right.\right.\\
 &\left.\left.\left.\left.+\alpha_{NP} -\text{$\delta $a}_\parallel
 +\text{$\delta $c}_0\right)\right)\right)+b_\parallel \left(d_0 \left(y \cos \left(\alpha_{SM}+\alpha_{NP} -\gamma -\text{$\delta $b}_\parallel+
 \text{$\delta $d}_0-\phi \right)-x \sin \left(\alpha_{SM}+\alpha_{NP}-\right.\right.\right.\right.\\
 &\left.\left.\left.\left.\gamma -\text{$\delta $b}_\parallel+\text{$\delta $d}_0-\phi \right)\right)+c_0 \left(y \cos \left(\alpha_{SM}+\alpha_{NP} -\text{$\delta $b}_\parallel+\text{$\delta $c}_0-\phi \right)-x \sin \left(\alpha_{SM}+\alpha_{NP} -\text{$\delta $b}_\parallel+\right.\right.\right.\right.\\
 &\left.\left.\left.\left.\text{$\delta $c}_0-\phi \right)\right)\right)\right)\\
\bar{Y}_{ii}^{\bar{\text{V}_1}\bar{\text{V}_2}}= &\frac{2}{r} \left(a_i \left(d_i \left(x \sin \left(\alpha_{SM}+\alpha_{NP} +\delta _i^{\text{ac}}-\gamma +\delta _\lambda^{\text{cd}}\right)+y \cos \left(\alpha_{SM}+\alpha_{NP} +\delta _i^{\text{ac}}-\gamma +\delta _\lambda^{\text{cd}}\right)\right)+c_i \left(x \sin \left(\right.\right.\right.\right.\\
&\left.\left.\left.\left.\alpha_{SM}+\alpha_{NP} +\delta _i^{\text{ac}}\right)+y \cos \left(\alpha_{SM}+\alpha_{NP}+\delta _i^{\text{ac}}\right)\right)\right)+b_i \left(c_i \left(x \sin \left(\alpha_{SM}+\alpha_{NP} -\delta _i^{\text{ab}}+\delta _i^{\text{ac}}-\phi \right)+\right.\right.\right.\\
&\left.\left.\left.y \cos \left(\alpha_{SM}+\alpha_{NP} -\delta _i^{\text{ab}}+\delta _i^{\text{ac}}-\phi \right)\right)+d_i \left(x \sin \left(\alpha_{SM}+\alpha_{NP} -\gamma +\delta _\lambda^{\text{cd}}-\phi \right)+y \cos \left(\alpha_{SM}+\right.\right.\right.\right.\\
&\left.\left.\left.\left.\alpha_{NP} -\gamma +\delta _\lambda^{\text{cd}}-\phi \right)\right)\right)\right)\\
\bar{Y}_{\perp\perp}^{\bar{\text{V}_1}\bar{\text{V}_2}}= &-\frac{2}{r} \left(a_\perp \left(d_\perp \left(x \sin \left(\alpha_{SM}+\alpha_{NP} +\delta _\perp^{\text{ac}}-\gamma +\delta _\perp^{\text{cd}}\right)+y \cos \left(\alpha_{SM}+\alpha_{NP} +\delta _\perp^{\text{ac}}-\gamma +\delta _\perp^{\text{cd}}\right)\right)+c_\perp \left(x\right.\right.\right.\\
&\left.\left.\left.\sin \left(\alpha_{SM}+\alpha_{NP} +\delta _\perp^{\text{ac}}\right)+y \cos \left(\alpha_{SM}+\alpha_{NP}
 +\delta _\perp^{\text{ac}}\right)\right)\right)+b_\perp \left(c_\perp \left(x \sin \left(\alpha_{SM}+\alpha_{NP} -\delta _\perp^{\text{ab}}+\delta _\perp^{\text{ac}}-\right.\right.\right.\right.\\
&\left.\left.\left.\left. \phi \right)+y \cos \left(\alpha_{SM}+\alpha_{NP} -\delta _\perp^{\text{ab}}+\delta _\perp^{\text{ac}}-\phi \right)\right)+d_\perp \left(x \sin \left(\alpha_{SM}+\alpha_{NP} -\gamma +
 \delta _\perp^{\text{cd}}-\phi \right)+y \cos \left(\alpha_{SM}\right.\right.\right.\right.\\
 &\left.\left.\left.\left.+\alpha_{NP} -\gamma +\delta _\perp^{\text{cd}}-\phi \right)\right)\right)\right)\\
\bar{Y}_{i\perp}^{\bar{\text{V}_1}\bar{\text{V}_2}}= &\frac{2}{r} \left(b_\perp \left(d_i \left(y \sin \left(\Delta _i^a+\alpha_{SM}+\alpha_{NP} -\delta _\perp^{\text{ab}}+\delta _i^{\text{ac}}-\gamma +\delta _\lambda^{\text{cd}}-\phi \right)-x \cos \left(\Delta _i^a+\alpha_{SM}+\alpha_{NP} -\delta _\perp^{\text{ab}}+\right.\right.\right.\right.\\
&\left.\left.\left.\left.\delta _i^{\text{ac}}-\gamma +\delta _\lambda^{\text{cd}}-\phi \right)\right)+c_i \left(y \sin \left(\Delta _i^a+\alpha_{SM}+\alpha_{NP} -\delta _\perp^{\text{ab}}+\delta _i^{\text{ac}}-\phi \right)-x \cos \left(\Delta _i^a+\alpha_{SM}+\alpha_{NP} -\right.\right.\right.\right.\\
&\left.\left.\left.\left.\delta _\perp^{\text{ab}}+\delta _i^{\text{ac}}-\phi \right)\right)\right)+d_\perp \left(b_i \left(y \sin \left(-\Delta _i^a+\alpha_{SM}+\alpha_{NP} +\delta _i^{\text{ab}}+\delta _\perp^{\text{ac}}-\gamma +\delta _\perp^{\text{cd}}-\phi \right)-x \cos \left(-\Delta _i^a\right.\right.\right.\right.\\
&\left.\left.\left.\left.+\alpha_{SM}+\alpha_{NP} +\delta _i^{\text{ab}}+\delta _\perp^{\text{ac}}-\gamma +\delta _\perp^{\text{cd}}-\phi \right)\right)+a_i \left(y \sin \left(-\Delta _i^a+\alpha_{SM}+\alpha_{NP} +\delta _\perp^{\text{ac}}-\gamma +\delta _\perp^{\text{cd}}\right)-x \right.\right.\right.\\
&\left.\left.\left.\cos \left(-\Delta _i^a+\alpha_{SM}+\alpha_{NP} +\delta _\perp^{\text{ac}}-\gamma +\delta _\perp^{\text{cd}}\right)\right)\right)+c_\perp \left(b_i \left(y \sin \left(-\Delta _i^a+\alpha_{SM}+\alpha_{NP} -\delta _i^{\text{ab}}+\delta _\perp^{\text{ac}}-\phi \right)\right.\right.\right.\\
&\left.\left.\left.-x \cos \left(-\Delta _i^a+\alpha_{SM}+\alpha_{NP} -\delta _i^{\text{ab}}+\delta _\perp^{\text{ac}}-\phi \right)\right)+a_i \left(y \sin \left(-\Delta _i^a+\alpha_{SM}+\alpha_{NP} +\delta _\perp^{\text{ac}}\right)-x \cos \left(\right.\right.\right.\right.
&\left.\left.\left.\left.\alpha_{SM}+\alpha_{NP} +\text{$\delta $a}_i-\text{$\delta $c}_\perp\right)\right)\right)+a_\perp \left(d_i \left(y \sin \left(\Delta _i^a+\alpha_{SM}+\alpha_{NP} +\delta _i^{\text{ac}}-\gamma +\delta _\lambda^{\text{cd}}\right)-x \cos \left(\Delta _i^a+\right.\right.\right.\right.\\
\end{align*}
\begin{align*}
&\left.\left.\left.\left.\alpha_{SM}+\alpha_{NP} +\delta _i^{\text{ac}}-\gamma +\delta _\lambda^{\text{cd}}\right)\right)+c_i \left(y \sin \left(\Delta _i^a+\alpha_{SM}+\alpha_{NP} +\delta _i^{\text{ac}}\right)-x \cos \left(\Delta _i^a+\alpha_{SM}+\alpha_{NP} +\right.\right.\right.\right.\\
&\left.\left.\left.\left.\delta _i^{\text{ac}}\right)\right)\right)\right)\\
\bar{Y}_{0\parallel}^{\bar{\text{V}_1}\bar{\text{V}_2}}= &\frac{2}{r} \left(b_\parallel \left(d_0 \left(x \sin \left(\Delta _0^a-\Delta _\parallel^a+\alpha_{SM}+\alpha_{NP} -\delta _\parallel^{\text{ab}}+\delta _0^{\text{ac}}-\gamma +\delta _0^{\text{cd}}-\phi \right)+y \cos \left(\Delta _0^a-\Delta _\parallel^a+\alpha_{SM}\right.\right.\right.\right.\\
&\left.\left.\left.\left.+\alpha_{NP} -\delta _\parallel^{\text{ab}}+\delta _0^{\text{ac}}-\gamma +\delta _0^{\text{cd}}-\phi \right)\right)+c_0 \left(x \sin \left(\Delta _0^a-\Delta _\parallel^a+\alpha_{SM}+\alpha_{NP} -\delta _\parallel^{\text{ab}}+\delta _0^{\text{ac}}-\phi \right)+y\right.\right.\right.\\
&\left.\left.\left.\cos \left(\Delta _0^a-\Delta _\parallel^a+\alpha_{SM}+\alpha_{NP} -\delta _\parallel^{\text{ab}}+\delta _0^{\text{ac}}-\phi \right)\right)\right)+d_\parallel \left(b_0 \left(x \sin \left(-\Delta _0^a+\Delta _\parallel^a+\alpha_{SM}+\alpha_{NP} -\right.\right.\right.\right.\\
&\left.\left.\left.\left.\delta _0^{\text{ab}}-\delta _\parallel^{\text{ac}}-\gamma +\delta _\parallel^{\text{cd}}\right)+y \cos \left(-\Delta _0^a+\Delta _\parallel^a+\alpha_{SM}+\alpha_{NP} -\delta _0^{\text{ab}}-
\delta _\parallel^{\text{ac}}-\gamma +\delta _\parallel^{\text{cd}}\right)\right)+a_0 \left(x \sin \left(-\right.\right.\right.\right.\\
&\left.\left.\left.\left.\Delta _0^a+\Delta _\parallel^a+\alpha_{SM}+\alpha_{NP} -\delta _\parallel^{\text{ac}}-\gamma +\delta _\parallel^{\text{cd}}\right)+y \cos \left(-\Delta _0^a+\Delta _\parallel^a+\alpha_{SM}+\alpha_{NP} -\delta _\parallel^{\text{ac}}-\gamma +\delta _\parallel^{\text{cd}}\right)\right)\right)\right.\\
&\left.+c_\parallel \left(b_0 \left(x \sin \left(-\Delta _0^a+\Delta _\parallel^a+\alpha_{SM}+\alpha_{NP} -\delta _0^{\text{ab}}+\delta _\parallel^{\text{ac}}-\phi \right)+y \cos \left(-\Delta _0^a+\Delta _\parallel^a+\alpha_{SM}+\alpha_{NP} -\right.\right.\right.\right.\\
&\left.\left.\left.\left.\delta _0^{\text{ab}}+\delta _\parallel^{\text{ac}}-\phi \right)\right)+a_0 \left(x \sin \left(-\Delta _0^a+\Delta _\parallel^a+\alpha_{SM}+\alpha_{NP} +\delta _\parallel^{\text{ac}}\right)+y \cos \left(-\Delta _0^a+\Delta _\parallel^a+\alpha_{SM}+\right.\right.\right.\right.\\
&\left.\left.\left.\left.\alpha_{NP} +\delta _\parallel^{\text{ac}}\right)\right)\right)+a_\parallel \left(d_0 \left(x \sin \left(\Delta _0^a-\Delta _\parallel^a+\alpha_{SM}+\alpha_{NP} +\delta _0^{\text{ac}}-\gamma +\delta _0^{\text{cd}}\right)+y \cos \left(\Delta _0^a-\Delta _\parallel^a+\right.\right.\right.\right.\\
&\left.\left.\left.\left.\alpha_{SM}+\alpha_{NP} +\delta _0^{\text{ac}}-\gamma +\delta _0^{\text{cd}}\right)\right)+c_0 \left(x \sin \left(\Delta _0^a-\Delta _\parallel^a+\alpha_{SM}+\alpha_{NP} +\delta _\parallel^{\text{ac}}\right)+y \cos \left(\Delta _0^a-\Delta _\parallel^a+\right.\right.\right.\right.\\
&\left.\left.\left.\left.\alpha_{SM}+\alpha_{NP} +\delta _\parallel^{\text{ac}}\right)\right)\right)\right)\\
\bar{Y}_{ii}^{\text{V}_1\text{V}_2}= &\frac{2}{r} \left(a_i \left(d_i \left(x \sin \left(\alpha_{SM}+\alpha_{NP} -\delta _i^{\text{ac}}-\gamma -\delta _\lambda^{\text{cd}}\right)+y \cos \left(\alpha_{SM}+\alpha_{NP} -\delta _i^{\text{ac}}-\gamma -\delta _\lambda^{\text{cd}}\right)\right)+c_i \left(x \sin \left(\right.\right.\right.\right.\\
&\left.\left.\left.\left.\alpha_{SM}+\alpha_{NP} -\delta _i^{\text{ac}}\right)+y \cos \left(\alpha_{SM}+\alpha_{NP} -\delta _i^{\text{ac}}\right)\right)\right)+b_i \left(d_i \left(x \sin \left(\alpha_{SM}+\alpha_{NP} +\delta _i^{\text{ab}}-\delta _i^{\text{ac}}-\gamma -\right.\right.\right.\right.\\
&\left.\left.\left.\left.\delta _\lambda^{\text{cd}}-\phi \right)+y \cos \left(\alpha_{SM}+\alpha_{NP} +\delta _i^{\text{ab}}-\delta _i^{\text{ac}}-\gamma -\delta _\lambda^{\text{cd}}-\phi \right)\right)+c_i \left(x \sin \left(\alpha_{SM}+\alpha_{NP} +\delta _i^{\text{ab}}-\delta _i^{\text{ac}}-\right.\right.\right.\right.\\
&\left.\left.\left.\left.\phi \right)+y \cos \left(\alpha_{SM}+\alpha_{NP} +\delta _i^{\text{ab}}-\delta _i^{\text{ac}}-\phi \right)\right)\right)\right)\\
\bar{Y}_{\perp\perp}^{\text{V}_1\text{V}_2}= &-\frac{2}{r} \left(a_\perp \left(d_\perp \left(x \sin \left(\alpha_{SM}+\alpha_{NP} -\delta _\perp^{\text{ac}}-\gamma -\delta _\perp^{\text{cd}}\right)+y \cos \left(\alpha_{SM}+\alpha_{NP} -\delta _\perp^{\text{ac}}-\gamma -\delta _\perp^{\text{cd}}\right)\right)+c_\perp \left(\right.\right.\right.\\
&\left.\left.\left.x \sin \left(\alpha_{SM}+\alpha_{NP} -\delta _\perp^{\text{ac}}\right)+y \cos \left(\alpha_{SM}+\alpha_{NP} -\delta _\perp^{\text{ac}}\right)\right)\right)+b_\perp \left(d_\perp \left(x \sin \left(\alpha_{SM}+\alpha_{NP} +\delta _\perp^{\text{ab}}-\delta _\perp^{\text{ac}}\right.\right.\right.\right.\\
&\left.\left.\left.\left.-\gamma -\delta _\perp^{\text{cd}}-\phi \right)+y \cos \left(\alpha_{SM}+\alpha_{NP} +\delta _\perp^{\text{ab}}-\delta _\perp^{\text{ac}}-\gamma -\delta _\perp^{\text{cd}}-\phi \right)\right)+c_\perp \left(x \sin \left(\alpha_{SM}+\alpha_{NP} +\delta _\perp^{\text{ab}}\right.\right.\right.\right.\\
&\left.\left.\left.\left.-\delta _\perp^{\text{ac}}-\phi \right)+y \cos \left(\alpha_{SM}+\alpha_{NP} +\delta _\perp^{\text{ab}}-\delta _\perp^{\text{ac}}-\phi \right)\right)\right)\right)\\
\bar{Y}_{i\perp}^{\text{V}_1\text{V}_2}= &\frac{2}{r} \left(b_\perp \left(d_i \left(y \sin \left(-\Delta _i^a+\alpha_{SM}+\alpha_{NP} +\delta _\perp^{\text{ab}}-\delta _i^{\text{ac}}-\gamma -\delta _\lambda^{\text{cd}}-\phi \right)-x \cos \left(-\Delta _i^a+\alpha_{SM}+\alpha_{NP} +\right.\right.\right.\right.\\
&\left.\left.\left.\left.\delta _\perp^{\text{ab}}-\delta _i^{\text{ac}}-\gamma -\delta _\lambda^{\text{cd}}-\phi \right)\right)+c_i \left(y \sin \left(-\Delta _i^a+\alpha_{SM}+\alpha_{NP} +\delta _\perp^{\text{ab}}-\delta _i^{\text{ac}}-\phi \right)-x \cos \left(-\Delta _i^a+\alpha_{SM}\right.\right.\right.\right.\\
&\left.\left.\left.\left.+\alpha_{NP} +\delta _\perp^{\text{ab}}-\delta _i^{\text{ac}}-\phi \right)\right)\right)+d_\perp \left(b_i \left(y \sin \left(\Delta _i^a+\alpha_{SM}+\alpha_{NP} -\delta _i^{\text{ab}}-\delta _\perp^{\text{ac}}-\gamma -\delta _\perp^{\text{cd}}-\phi \right)-x \cos \right.\right.\right.\\
&\left.\left.\left.\left(\Delta _i^a+\alpha_{SM}+\alpha_{NP} -\delta _i^{\text{ab}}-\delta _\perp^{\text{ac}}-\gamma -\delta _\perp^{\text{cd}}-\phi \right)\right)+a_i \left(y \sin \left(\Delta _i^a+\alpha_{SM}+\alpha_{NP} -\delta _\perp^{\text{ac}}-\gamma -\delta _\perp^{\text{cd}}\right) \right.\right.\right.\\
&\left.\left.\left.-x \cos \left(\Delta _i^a+\alpha_{SM}+\alpha_{NP} -\delta _\perp^{\text{ac}}-\gamma -\delta _\perp^{\text{cd}}\right)\right)\right)+c_\perp \left(b_i \left(y \sin\left(\Delta _i^a+\alpha_{SM}+\alpha_{NP} +\delta _i^{\text{ab}}-\delta _\perp^{\text{ac}}-\phi \right)\right.\right.\right.\\
&\left.\left.\left.-x \cos \left(\Delta _i^a+\alpha_{SM}+\alpha_{NP} +\delta _i^{\text{ab}}-\delta _\perp^{\text{ac}}-\phi \right)\right)+a_i \left(y \sin \left(\Delta _i^a+\alpha_{SM}+\alpha_{NP} -\delta _\perp^{\text{ac}}\right)-x \cos \left(\Delta _i^a+\right.\right.\right.\right.\\
&\left.\left.\left.\left.\alpha_{SM}+\alpha_{NP} -\delta _\perp^{\text{ac}}\right)\right)\right)+a_\perp \left(d_i \left(y \sin \left(-\Delta _i^a+\alpha_{SM}+\alpha_{NP} -\delta _i^{\text{ac}}-\gamma -\delta _\lambda^{\text{cd}}\right)-x \cos \left(-\Delta _i^a+\alpha_{SM}\right.\right.\right.\right.\\
\end{align*}
\begin{align*}
&\left.\left.\left.\left.-\gamma +\alpha_{NP} -\delta _i^{\text{ac}}-\delta _\lambda^{\text{cd}}\right)\right)+c_i \left(y \sin \left(-\Delta _i^a+\alpha_{SM}+\alpha_{NP} -\delta _i^{\text{ac}}\right)-x \cos \left(-\Delta _i^a+
\alpha_{SM}+\alpha_{NP} -\right.\right.\right.\right.\\
&\left.\left.\left.\left.\delta _i^{\text{ac}}\right)\right)\right)\right)\\
\bar{Y}_{0\parallel}^{\text{V}_1\text{V}_2}= &\frac{2}{r} \left(b_\parallel \left(d_0 \left(x \sin \left(-\Delta _0^a+\Delta _\parallel^a+\alpha_{SM}+\alpha_{NP} +\delta _\parallel^{\text{ab}}-\delta _0^{\text{ac}}-\gamma -\delta _0^{\text{cd}}-\phi \right)+y \cos \left(-\Delta _0^a+\Delta _\parallel^a+\right.\right.\right.\right.\\
&\left.\left.\left.\left.\alpha_{SM}+\alpha_{NP} +\delta _\parallel^{\text{ab}}-\delta _0^{\text{ac}}-\gamma -\delta _0^{\text{cd}}-\phi \right)\right)+c_0 \left(x \sin \left(-\Delta _0^a+\Delta _\parallel^a+\alpha_{SM}+\alpha_{NP} +\delta _\parallel^{\text{ab}}-\delta _0^{\text{ac}}-\phi \right.\right.\right.\right.\\
&\left.\left.\left.\left.\right)+y \cos \left(-\Delta _0^a+\Delta _\parallel^a+\alpha_{SM}+\alpha_{NP} +\delta _\parallel^{\text{ab}}-\delta _0^{\text{ac}}-\phi \right)\right)\right)+d_\parallel \left(b_0 \left(x \sin \left(\Delta _0^a-\Delta _\parallel^a+\alpha_{SM}+\right.\right.\right.\right.\\
&\left.\left.\left.\left.\alpha_{NP} +\delta _0^{\text{ab}}-\delta _\parallel^{\text{ac}}-\gamma -\delta _\parallel^{\text{cd}}-\phi \right)+y \cos \left(\Delta _0^a-\Delta _\parallel^a+\alpha_{SM}+\alpha_{NP} +
\delta _0^{\text{ab}}-\delta _\parallel^{\text{ac}}-\gamma -\delta _\parallel^{\text{cd}}-\phi \right)\right)\right.\right.\\
&\left.\left.+a_0 \left(x \sin \left(\Delta _0^a-\Delta _\parallel^a+\alpha_{SM}+\alpha_{NP} -\delta _\parallel^{\text{ac}}-\gamma -\delta _\parallel^{\text{cd}}\right)+y \cos \left(\Delta _0^a-\Delta _\parallel^a+\alpha_{SM}+\alpha_{NP}-\delta _\parallel^{\text{ac}}-\right.\right.\right.\right.\\
&\left.\left.\left.\left.\gamma-\delta _\parallel^{\text{cd}}\right)\right)\right)+c_\parallel \left(b_0 \left(y \cos \left(\Delta _0^a-\Delta _\parallel^a+\alpha_{SM}+\alpha_{NP} +\delta _0^{\text{ab}}-\delta _\parallel^{\text{ac}}-\phi \right)+x \sin \left(\alpha_{SM}+\alpha_{NP} \right.\right.\right.\right.\\
&\left.\left.\left.\left.+\delta _0^{\text{ab}}-\delta _\parallel^{\text{ac}}-\phi \right)\right)+a_0 \left(x \sin \left(\Delta _0^a-\Delta _\parallel^a+\alpha_{SM}+\alpha_{NP} -\delta _\parallel^{\text{ac}}\right)+y \cos \left(\Delta _0^a-\Delta _\parallel^a+\alpha_{SM}+\alpha_{NP} \right.\right.\right.\right.\\
&\left.\left.\left.\left.-\delta _\parallel^{\text{ac}}\right)\right)\right)+a_\parallel \left(d_0 \left(x \sin \left(-\Delta _0^a+\Delta _\parallel^a+\alpha_{SM}+\alpha_{NP} -\delta _0^{\text{ac}}-\gamma -\delta _0^{\text{cd}}\right)+y \cos \left(-\Delta _0^a+\Delta _\parallel^a+\alpha_{SM}\right.\right.\right.\right.\\
&\left.\left.\left.\left.+\alpha_{NP} -\delta _0^{\text{ac}}-\gamma -\delta _0^{\text{cd}}\right)\right)+c_0 \left(x \sin \left(-\Delta _0^a+\Delta _\parallel^a+\alpha_{SM}+\alpha_{NP} -\delta _0^{\text{ac}}\right)+y \cos \left(-\Delta _0^a+\Delta _\parallel^a+\right.\right.\right.\right.\\
&\left.\left.\left.\left.\alpha_{SM}+\alpha_{NP} -\delta _0^{\text{ac}}\right)\right)\right)\right)\\
 Y_{ii}^{\bar{\text{V}_1}\bar{\text{V}_2} }= &2 r \left(a_i \left(d_i \left(y \cos \left(\alpha_{SM}+\alpha_{NP} +\delta _i^{\text{ac}}-\gamma +\delta _\lambda^{\text{cd}}\right)-x \sin \left(\alpha_{SM}+\alpha_{NP} +\delta _i^{\text{ac}}-\gamma +\delta _\lambda^{\text{cd}}\right)\right)+c_i \left(y \cos\right.\right.\right.\\
 &\left.\left.\left.\left(\alpha_{SM}+\alpha_{NP} +\delta _i^{\text{ac}}\right)-x \sin \left(\alpha_{SM}+\alpha_{NP} +\delta _i^{\text{ac}}\right)\right)\right)+b_i \left(c_i \left(y \cos \left(\alpha_{SM}+\alpha_{NP} -\delta _i^{\text{ab}}+\delta _i^{\text{ac}}-\phi \right)-\right.\right.\right.\\
 &\left.\left.\left.x \sin \left(\alpha_{SM}+\alpha_{NP} -\delta _i^{\text{ab}}+\delta _i^{\text{ac}}-\phi \right)\right)+d_i \left(y \cos \left(\alpha_{SM}+\alpha_{NP} -\gamma +\delta _\lambda^{\text{cd}}-\phi \right)-x \sin \left(\alpha_{SM}+\right.\right.\right.\right.\\
 &\left.\left.\left.\left.\alpha_{NP} -\gamma +\delta _\lambda^{\text{cd}}-\phi \right)\right)\right)\right)\\
Y_{\perp\perp}^{\bar{\text{V}_1}\bar{\text{V}_2} }= &-2 r \left(a_\perp \left(d_\perp \left(y \cos \left(\alpha_{SM}+\alpha_{NP} +\delta _\perp^{\text{ac}}-\gamma +\delta _\perp^{\text{cd}}\right)-x \sin \left(\alpha_{SM}+\alpha_{NP} +\delta _\perp^{\text{ac}}-\gamma +\delta _\perp^{\text{cd}}\right)\right)+c_\perp \left(\right.\right.\right.\\
&\left.\left.\left.y \cos \left(\alpha_{SM}+\alpha_{NP} +\delta _\perp^{\text{ac}}\right)-x \sin \left(\alpha_{SM}+\alpha_{NP} +\delta _\perp^{\text{ac}}\right)\right)\right)+b_\perp \left(c_\perp \left(y \cos \left(\alpha_{SM}+\alpha_{NP} -\delta _\perp^{\text{ab}}+\delta _\perp^{\text{ac}} \right.\right.\right.\right.\\
&\left.\left.\left.\left.-\phi \right)-x \sin \left(\alpha_{SM}+\alpha_{NP} -\delta _\perp^{\text{ab}}+\delta _\perp^{\text{ac}}-\phi \right)\right)+d_\perp \left(y \cos \left(\alpha_{SM}+\alpha_{NP} -\gamma +\delta _\perp^{\text{cd}}-\phi \right)-x \sin \left(\right.\right.\right.\right.\\
&\left.\left.\left.\left.\alpha_{SM}+\alpha_{NP} -\gamma +\delta _\perp^{\text{cd}}-\phi \right)\right)\right)\right)\\
Y_{i\perp}^{\bar{\text{V}_1}\bar{\text{V}_2} }= &2 r \left(b_\perp \left(d_i \left(x \cos \left(\Delta _i^a+\alpha_{SM}+\alpha_{NP} -\delta _\perp^{\text{ab}}+\delta _i^{\text{ac}}-\gamma +\delta _\lambda^{\text{cd}}-\phi \right)+y \sin \left(\Delta _i^a+\alpha_{SM}+\alpha_{NP} -\delta _\perp^{\text{ab}}\right.\right.\right.\right.\\
&\left.\left.\left.\left.+\delta _i^{\text{ac}}-\gamma +\delta _\lambda^{\text{cd}}-\phi \right)\right)+c_i \left(x \cos \left(\Delta _i^a+\alpha_{SM}+\alpha_{NP} -\delta _\perp^{\text{ab}}+\delta _i^{\text{ac}}-\phi \right)+y \sin \left(\Delta _i^a+\alpha_{SM}+\alpha_{NP} \right.\right.\right.\right.\\
&\left.\left.\left.\left.-\delta _\perp^{\text{ab}}+\delta _i^{\text{ac}}-\phi \right)\right)\right)+d_\perp \left(b_i \left(x \cos \left(-\Delta _i^a+\alpha_{SM}+\alpha_{NP} +\delta _i^{\text{ab}}+\delta _\perp^{\text{ac}}-\gamma +\delta _\perp^{\text{cd}}-\phi \right)+y \sin \left(-\Delta _i^a\right.\right.\right.\right.\\
&\left.\left.\left.\left.+\alpha_{SM}+\alpha_{NP} +\delta _i^{\text{ab}}+\delta _\perp^{\text{ac}}-\gamma +\delta _\perp^{\text{cd}}-\phi \right)\right)+a_i \left(x \cos \left(-\Delta _i^a+\alpha_{SM}+\alpha_{NP} +\delta _\perp^{\text{ac}}-\gamma +\delta _\perp^{\text{cd}}\right)+y \right.\right.\right.\\
&\left.\left.\left.\sin \left(-\Delta _i^a+\alpha_{SM}+\alpha_{NP} +\delta _\perp^{\text{ac}}-\gamma +\delta _\perp^{\text{cd}}\right)\right)\right)+c_\perp \left(b_i \left(x \cos \left(-\Delta _i^a+\alpha_{SM}+\alpha_{NP} -\delta _i^{\text{ab}}+\delta _\perp^{\text{ac}}-\phi\right.\right.\right.\right.\\
 &\left.\left.\left.\left. \right)+y \sin \left(-\Delta _i^a+\alpha_{SM}+\alpha_{NP} -\delta _i^{\text{ab}}+\delta _\perp^{\text{ac}}-\phi \right)\right)+a_i \left(x \cos \left(-\Delta _i^a+\alpha_{SM}+\alpha_{NP} +\delta _\perp^{\text{ac}}\right)+y \sin \left(\right.\right.\right.\right.\\
 &\left.\left.\left.\left.-\Delta _i^a+\alpha_{SM}+\alpha_{NP} +\delta _\perp^{\text{ac}}\right)\right)\right)+a_\perp \left(d_i \left(x \cos \left(\Delta _i^a+\alpha_{SM}+\alpha_{NP} +\delta _i^{\text{ac}}-\gamma +\delta _\lambda^{\text{cd}}\right)+y \sin \left(\Delta _i^a+\right.\right.\right.\right.\\
 \end{align*}
\begin{align*}
&\left.\left.\left.\left.\alpha_{SM}+\alpha_{NP} +\delta _i^{\text{ac}}-\gamma +\delta _\lambda^{\text{cd}}\right)\right)+c_i \left(x \cos \left(\Delta _i^a+\alpha_{SM}+\alpha_{NP}+\delta _i^{\text{ac}}\right)+y \sin \left(\Delta _i^a+\alpha_{SM}+\alpha_{NP}\right.\right.\right.\right.\\
 &\left.\left.\left.\left.+\delta _i^{\text{ac}}\right)\right)\right)\right)\\
 Y_{0\parallel}^{\bar{\text{V}_1}\bar{\text{V}_2}}= &2 r \left(b_\parallel \left(d_0 \left(y \cos \left(\Delta _0^a-\Delta _\parallel^a+\alpha_{SM}+\alpha_{NP} -\delta _\parallel^{\text{ab}}+\delta _0^{\text{ac}}-\gamma +\delta _0^{\text{cd}}-\phi \right)-x \sin \left(\Delta _0^a-\Delta _\parallel^a+\alpha_{SM}\right.\right.\right.\right.\\
 &\left.\left.\left.\left.+\alpha_{NP} -\delta _\parallel^{\text{ab}}+\delta _0^{\text{ac}}-\gamma +\delta _0^{\text{cd}}-\phi \right)\right)+c_0 \left(y \cos \left(\Delta _0^a-\Delta _\parallel^a+\alpha_{SM}+\alpha_{NP} -\delta _\parallel^{\text{ab}}+\delta _0^{\text{ac}}-\phi \right)-x \right.\right.\right.\\
 &\left.\left.\left.\sin \left(\Delta _0^a-\Delta _\parallel^a+\alpha_{SM}+\alpha_{NP} -\delta _\parallel^{\text{ab}}+\delta _0^{\text{ac}}-\phi \right)\right)\right)+d_\parallel \left(b_0 \left(y \cos \left(-\Delta _0^a+\Delta _\parallel^a+\alpha_{SM}+\alpha_{NP} -\right.\right.\right.\right.\\
 &\left.\left.\left.\left.\delta _0^{\text{ab}}-\delta _\parallel^{\text{ac}}-\gamma +\delta _\parallel^{\text{cd}}\right)-x \sin \left(-\Delta _0^a+\Delta _\parallel^a+\alpha_{SM}+\alpha_{NP} -\delta _0^{\text{ab}}
 -\delta _\parallel^{\text{ac}}-\gamma +\delta _\parallel^{\text{cd}}\right)\right)+a_0 \left(y \cos \left(-\right.\right.\right.\right.\\
 &\left.\left.\left.\left.\Delta _0^a+\Delta _\parallel^a+\alpha_{SM}+\alpha_{NP} -\delta _\parallel^{\text{ac}}-\gamma +\delta _\parallel^{\text{cd}}\right)-x \sin \left(-\Delta _0^a+\Delta _\parallel^a+\alpha_{SM}+\alpha_{NP} -\delta _\parallel^{\text{ac}}-\gamma +\delta _\parallel^{\text{cd}}\right)\right)\right)\right.\\
 &\left.+c_\parallel \left(b_0 \left(y \cos \left(-\Delta _0^a+\Delta _\parallel^a+\alpha_{SM}+\alpha_{NP} -\delta _0^{\text{ab}}+\delta _\parallel^{\text{ac}}-\phi \right)-x \sin \left(-\Delta _0^a+\Delta _\parallel^a+\alpha_{SM}+\alpha_{NP} -\right.\right.\right.\right.\\
 &\left.\left.\left.\left.\delta _0^{\text{ab}}+\delta _\parallel^{\text{ac}}-\phi \right)\right)+a_0 \left(y \cos \left(-\Delta _0^a+\Delta _\parallel^a+\alpha_{SM}+\alpha_{NP} +\delta _\parallel^{\text{ac}}\right)-x \sin \left(-\Delta _0^a+\Delta _\parallel^a+\alpha_{SM}+\right.\right.\right.\right.\\
 &\left.\left.\left.\left.\alpha_{NP} +\delta _\parallel^{\text{ac}}\right)\right)\right)+a_\parallel \left(d_0 \left(y \cos \left(\Delta _0^a-\Delta _\parallel^a+\alpha_{SM}+\alpha_{NP} +\delta _0^{\text{ac}}-\gamma +\delta _0^{\text{cd}}\right)-x \sin \left(\Delta _0^a-\Delta _\parallel^a+\right.\right.\right.\right.\\
 &\left.\left.\left.\left.\alpha_{SM}+\alpha_{NP} +\delta _0^{\text{ac}}-\gamma +\delta _0^{\text{cd}}\right)\right)+c_0 \left(y \cos\left(\Delta _0^a-\Delta _\parallel^a+\alpha_{SM}+\alpha_{NP} +\delta _\parallel^{\text{ac}}\right)-x \sin \left(\Delta _0^a-\Delta _\parallel^a+\right.\right.\right.\right.\\
 &\left.\left.\left.\left.\alpha_{SM}+\alpha_{NP} +\delta _\parallel^{\text{ac}}\right)\right)\right)\right)
\end{align*}

%\bibliography{cite-paper}


\end{document}